\documentclass[apj]{emulateapj}

\usepackage{natbib}
\usepackage{graphicx}
\bibpunct{(}{)}{;}{a}{}{,}

\newcommand\bcd{{\rm BCD}}
\newcommand\qbcd{{\rm QBCD}}
\newcommand\miles{{\rm MILES~SSP}}
\newcommand\sfr{{\rm SFR}}
\newcommand\paperi{\rm Paper~I}
\epsscale{0.5}

\begin{document}
%
\title{
Search for blue compact dwarf galaxies during quiescence II:\\ 
	metallicities of gas and stars, ages, and star-formation rates}
\author{
	J.~S\'anchez~Almeida\altaffilmark{1} 
	J.~A.~L.~Aguerri\altaffilmark{1},  \\
	C.~Mu\~noz-Tu\~n\'on\altaffilmark{1},  
	 A.~Vazdekis\altaffilmark{1}
          }

\altaffiltext{1}{Instituto de Astrof\'\i sica de Canarias, E-38205 La Laguna, Tenerife, Spain}
 
            \email{jos@iac.es, jalfonso@iac.es, cmt@iac.es, vazdekis@iac.es}

\begin{abstract}
We examine the metallicity and age of a large set of 
SDSS/DR6 galaxies that may be Blue Compact Dwarf (\bcd ) 
galaxies during quiescence (\qbcd s).
The individual spectra are first classified and then
averaged to reduce noise.
The metallicity inferred 
from emission lines (tracing ionized gas) exceeds 
by $\sim$0.35~dex the 
metallicity inferred from absorption lines (tracing stars). 
Such a small  difference  is significant 
according to our error budget estimate.
The same procedure was applied to a reference 
sample of \bcd s, and in this case the two
metallicities agree,  being also consistent 
with the stellar metallicity in \qbcd s.  
Chemical evolution models indicate that
the gas metallicity of \qbcd s  is  too high to be 
representative of the galaxy as a  whole, but it can   
represent a small fraction of the galactic gas, 
self enriched by previous starbursts. 
The luminosity weighted stellar age 
of \qbcd s  spans the whole range between 1 and 
10~Gyr,
whereas it is always smaller than 1~Gyr for \bcd s.
Our stellar ages and metallicities rely on
a single stellar population
spectrum fitting procedure, which we have specifically developed 
for this work using the stellar library MILES.
%
%
\end{abstract}



\shorttitle{BCD Galaxies During Quiescence II. Metallicities and ages}
\shortauthors{S\'anchez Almeida et al.}

%


\section{Introduction}\label{introduction}

In the hierarchical 
picture of galaxy formation, large galaxies  arise through the 
assembly of smaller aggregates \citep[e.g.,][]{whi91,die07},
and metal-deficient dwarf galaxies are possibly the closest 
examples we can find of the elementary primordial units from 
which galaxies assembled.
In the downsizing paradigm, large galaxies form metals 
early on  \citep[e.g.,][]{cow96,kau03b},
and only low metallicity dwarfs  may still keep a 
fossil record of the pristine  Inter-Stellar Medium 
(ISM). For one reason or another, 
Blue Compact Dwarf (\bcd ) galaxies seem to probe early 
phases of the Universe. They combine the two 
required ingredients, i.e., they are dwarfs having
record-braking low metallicities 
\citep[e.g.][]{kun00,izo05}. 
\bcd\ galaxies 
have been used to constrain,
e.g., the  properties
of the first (Pop~III) stars that polluted the primordial 
ISM at the time of galaxy formation \citep[e.g.,][]{bro04}, 
or the primordial He abundance inherited from big-bang nucleosynthesis 
\citep[e.g.,][]{izo07b}. 

%
%
%
%

\bcd\ galaxies have been extensively
studied during the last 35 years \citep[e.g.][]{sar70,
sea72,sil87,dav88,pap96,tel97,cai01,gil03,gil05,amo07}.
However, the way they grow up and evolve
remains unknown. The
intense starburst that characterizes 
\bcd s lasts only a few Myr 
\citep[e.g.,][]{mas99,thu91}. 
BCDs seem to be undergoing a transient but we 
ignore
how they reach such state, and 
what happens to them afterwards.
Consequently, identifying precursors and
descendants of \bcd s would be a major 
breakthrough in deciphering the nature and
the functioning of these special galaxies. It will facilitate
using them as reliable tools in cosmological studies.

Evolutionary connections between different 
dwarf galaxies and \bcd s have been both proposed 
and questioned in the literature 
\citep[e.g.,][]{sea72,sil87,dav88,pap96,tel97,gil05,del07,amo07b}.
In an attempt to complement these efforts,
we carried out a search for galaxies that may be
\bcd\ during the periods where the major starburst
is gone, i.e., quiescent~\bcd s or, for short, 
\qbcd s \citep[][hereafter \paperi]{san08}.
We addressed the issue from a new perspective.
Most  \bcd s show a red low surface brightness
component which should exist before the present 
starburst sets in and should remain once the
starburst fades away \citep[][]{loo86,pap96b}.
By carefully removing  
the starburst,
this underlying component has been 
studied and characterized
in the literature  \citep[e.g.,][]{noe03,cao05,amo07,amo07b}.
We searched the SDSS/DR6~database for isolated
galaxies
with the luminosity, color,
surface brightness, and concentration 
characteristic of the low surface brightness
component underlying the \bcd s (\paperi).
Assuming that the underlying low surface
brightness galaxy remains  unaltered
after each starburst
exhaustion, the targets thus selected 
could be
\qbcd s. 
The search yielded some~21500 \qbcd\ candidates, 
with properties pointing out that they may be
indeed pre or post
\bcd s. In particular, they have the 
same luminosity function as the \bcd s, although 
they are thirty times more 
numerous. The results suggested
an evolutive 
sequence where \bcd s undergo
many short starburst phases during their 
lifetimes, as proposed long ago by
\citet{sea72}. In between bursts, the galaxies
show up as \qbcd s in a low activity
state of various degrees 
lasting thirty times longer than the 
bursts. Statistically, \qbcd s should undergo
a \bcd\ phase every 300~Myr and lasting some
10~Myr.
This sequence 
of \bcd\ and 
\qbcd\ phases
can be maintained during several
Hubble times, and the most active \qbcd s are indeed
\bcd s. \paperi\ carries out the differential comparison
with \bcd s by selecting the sample of \bcd s 
also from SDSS/DR6, and employing
the same procedures used to
retrieve the \qbcd s.

In spite of all these agreeable features, the
evolutive link between \bcd s and \qbcd s
presents an important difficulty
posed in \paperi .
The \qbcd\ oxygen abundance
was estimated to be 
0.35~dex systematically larger than 
the oxygen abundance of the \bcd s. This makes the
role of \qbcd s as \bcd\ precursors questionable
since starbursts increase metallicity, and 
the putative precursors (\qbcd s) should have lower 
metallicity than their descendants (\bcd s).
\paperi\ offered
a few alternatives to clear out the difficulty,
most of which were related with the infall 
of metal poor gas before the starburst sets in. 
In addition,  we speculated 
that the metallicity assigned to the \qbcd s
may be biased, with the
{\em true} \qbcd\ metallicities much lower than
the observed ones, and close to the
observed \bcd\ metallicities.  
We derive the oxygen abundance from emission 
lines produced in H~{\sc ii} regions, which trace 
the ISM in those places now
going through a star-formation 
episode. In the case of \qbcd s, the 
star formation rate is quite small 
($< 0.1\,M_\odot$\,y$^{-1}$ even for the brightest
ones; \S~\ref{chemical}), therefore the volume 
of galactic gas sampled
by the measurement is very small too.
The question arises as to whether the abundance 
of this gas is representative of the total galactic
gas. If it is not, then it could explain a 
false overabundance of oxygen
in our \qbcd\ candidates. The sampled gas  
may not be properly mixed up with the galactic
ISM and, therefore, be self metal enriched by successive 
starbursts.

The possibility that the metallicity deduced
from emission lines may be contaminated by recent
starbursts has been previously mentioned in the
literature \citep[e.g.,][]{kun86,thu04,dal07}.
The mixing of the ISM is a slow process, which leaves 
behind a patchy medium 
\citep[e.g.,][]{ten96,dea02}.

The present work was originally meant at
testing the main conjecture in \paperi , namely, that the emission 
line derived metallicity overestimates the 
{\em true} average metallicity of the \qbcd\ gas.
If so, it should be significantly larger than the
metallicity of other galactic components, in particular,
the metallicity of the stars. This seems 
to be the case (\S~\ref{metal_vs_metal}), but  
in the way of 
working it out, 
several other properties of 
\qbcd s (and \bcd s) have emerged. 
These results are described here
in fairly broad terms, keeping in mind
their potential interest outside the specific original 
motivation of the work. 
In particular, \qbcd\ galaxies are quite 
common (one out of each three 
local dwarfs; \paperi), so that
their properties may also be representative of the 
whole class of dwarf galaxies.

The paper is organized as follows:
\S~\ref{spectra} summarizes the main
observational properties of the
Sloan Digital Sky Survey/Data Release 6  
(SDSS/DR6) spectra used in our analysis.
\S~\ref{class} explains the classification
of spectra before averaging them
out to improve the signal-to-noise ratio.
By fitting the observed \qbcd\ spectra with 
synthetic spectra, we assign ages and
metallicities to the stellar component 
of the galaxies
(\S~\ref{metalic}).
Gas metallicities are estimated in \S~\ref{gas_metal},
with their uncertainties critically assesses
in App.~\ref{appa}.
\S~\ref{metal_vs_metal} puts forward the
excess of gas metallicity with respect
to the metallicity of the stellar component.
Ages and stellar content
are analyzed in \S~\ref{ages_sect}.
Chemical evolution model galaxies able
to account for the differences between
stellar and nebular metallicities are
discussed in \S~\ref{chemical}, where we
also have to estimate Star Formation Rates (SFRs).
Finally, the main results and their
implications are discussed in \S~\ref{conclusions}.

\section{Data set: SDSS spectra}\label{spectra} 

We aim at assigning  
metallicities to the stellar 
component of the \qbcd\ candidates selected in 
\paperi . The original galaxies were chosen
within the SDSS spectroscopic catalog,
to have redshifts from which we derive absolute  
magnitudes 
(mean redshift 0.030, with a standard deviation of 0.014).
The present analysis of stellar
metallicities is based on these SDSS 
spectra. 
For the sake of comprehensiveness, their main 
characteristics are summarized here. 
A more detailled account can be found in \citet{sto02},
\citet{ade08}, and also in the 
SDSS~website ({\tt http://www.sdss.org/dr6}).

The SDSS spectrograph has two independent 
arms, with a dichroic separating
the blue beam and the red beam at 6150 \AA .
It  simultaneously renders a spectral range 
from 3800\,\AA\ to 9200\,\AA , with 
a spectral resolution between 1800
and 2200. The sampling is linear in 
logarithmic wavelength, with a mean
dispersion of 1.1\,\AA\,pix$^{-1}$ in the blue 
and 1.8\,\AA\,pix$^{-1}$ in the red. 
Repeated 15 min exposure spectra are 
integrated to yield a S/N per pixel $>4$  
when the apparent magnitude in the  
$g~$bandpass is 20.2.
The spectrograph is fed by 
fibers which subtend about 3\arcsec
on the sky. Most 
galaxies are larger than
this size, therefore, the spectra sample 
only their central parts 
(e.g., 89\%\ of the \qbcd\ galaxies have an effective
radius larger than half the fiber diameter).
We retrieve the 21493 \qbcd\ spectra and the
1609 \bcd\ spectra in FITS format from 
the SDSS Data Archive Server. All
spectra were re-sampled to a common
restframe wavelength scale that matches the 
spectral library used in 
\S~\ref{metalic} \citep{san06,cen07}.
We use linear interpolation to
oversample the original spectra
with a constant dispersion 
of 0.9\,\AA\,pix$^{-1}$.
The spectra were normalized
to the flux in the $g$~color filter
(effective wavelength $\simeq$ 4825\,\AA ),
a normalization factor that we compute from
each spectra using the transmission curve
downloaded from the SDSS website.

\section{Classification of galaxy spectra}\label{class}

\begin{figure}
\includegraphics[width=0.35\textwidth,angle=90]{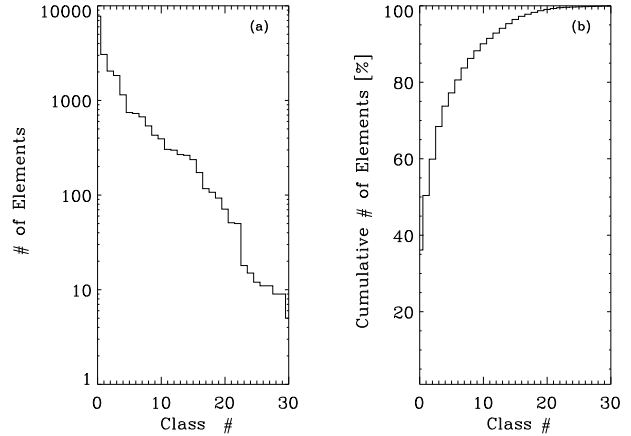}
\caption{(a) Histogram with the number of \qbcd\ 
	galaxies corresponding to each class.
	(b) Normalized cumulative histogram, i.e.,
	fraction of \qbcd\ galaxies from Class~0 to the
	each class~\#. It is given in percent. 
	Note that the first ten classes include 
	90~\% of the \qbcd s.  
}
\label{QBCDnumbers}
\end{figure}

\begin{figure*}
\includegraphics[width=0.7\textwidth,angle=90]{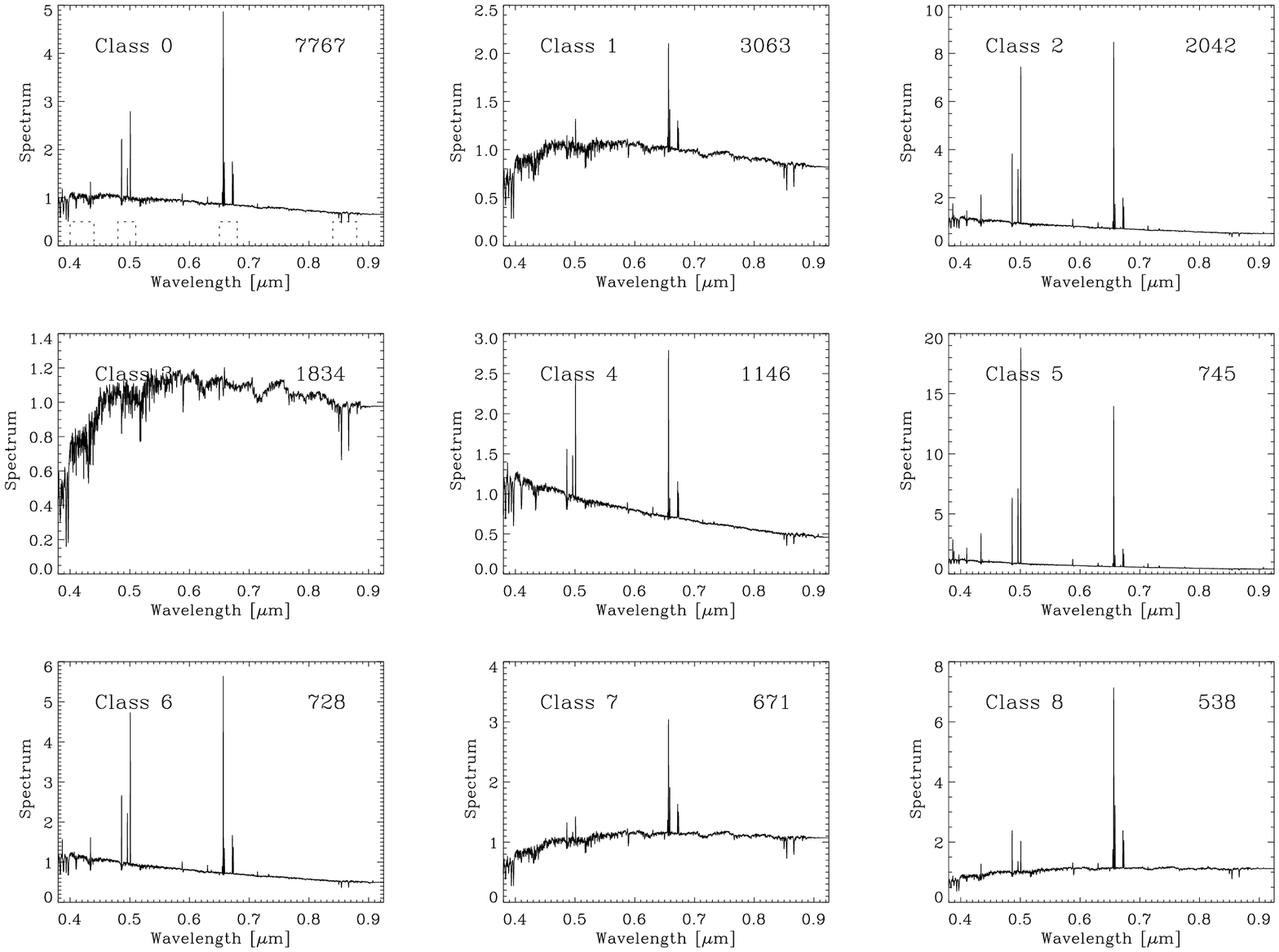}
\caption{Average spectra of the first nine most abundant
 \qbcd\ classes. Together with the
class identifier, the insets of the figures give the number of galaxies 
in the class. Wavelengths
are in $\mu {\rm m}$, and the range of ordinates is
different for the different plots.
All spectra are normalized to the flux in the 
$g$ filter. 
The dotted line shown together with Class~0 indicates
the band-passes used to classify spectra (i.e.,
those wavelengths were it is  
zero were disregarded for classification). 
}
\label{QBCDclasses}
\end{figure*}

\begin{figure*}
\includegraphics[width=0.7\textwidth,angle=90]{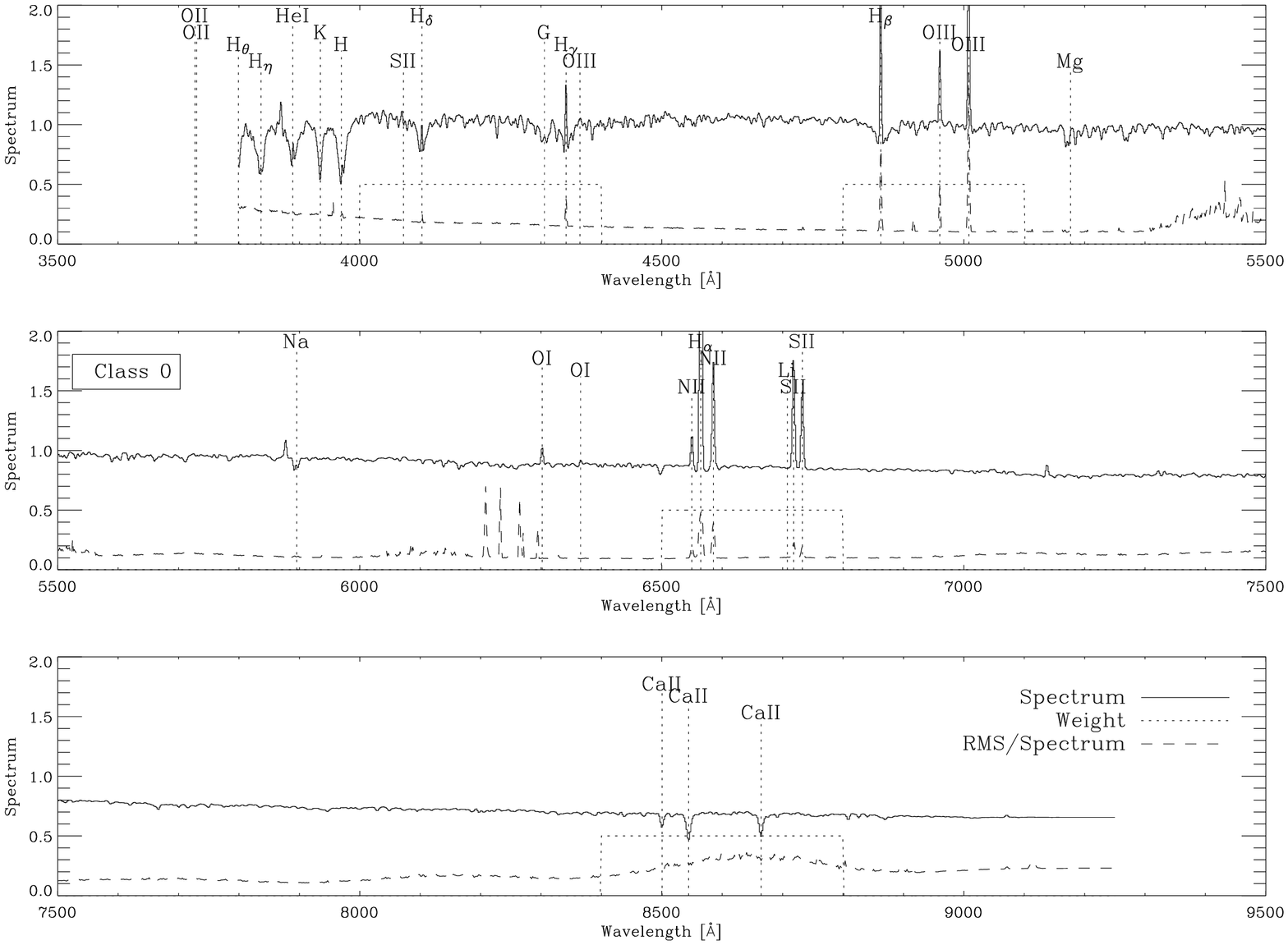}
\caption{
Expanded version of the spectrum of Class~0 in
Fig.~\ref{QBCDclasses}. In addition
to the spectrum itself (the solid line), the plot includes
the $rms$  fluctuations among all the 
spectra in the class divided by the mean spectrum
(the dashed line). The bandpasses used for classification
are shown as  dotted line. The plot also
labels several typical spectral features.
Wavelengths are given in \AA .
The large  fluctuations
between 6200\AA\  and 6300\AA\
are produced by telluric lines 
\citep[e.g., \S~4.2 in][]{gra09}.
Their presence does not affect our fits 
(see the residuals at these wavelengths in 
Figs.~\ref{fits}  and \ref{fitsb}). 
}
\label{classrms}
\end{figure*}

As we will discuss later on (\S~\ref{individual}), 
the S/N of the individual SDSS spectra is insufficient to 
estimate the metallicity of the stellar component. 
We improve the 
S/N ratio to acceptable levels by averaging similar 
spectra (a technique
often refereed to as {\em stacking}; 
see, e.g. \citealt{ell00}). 
Before averaging, the spectra have been
classified in alike sets using a cluster analysis algorithm. We
employ the simple {\em k-means clustering}
\citep[see, e.g., ][Chapter~5]{eve95}.
A number $k$ of template spectra are selected at random
from the full set. Each template spectrum is assumed to 
be a  cluster center, and each spectrum of the data 
set is assigned to the closest cluster center (closest
in a least squares sense). Once all spectra in the dataset
have been classified, the cluster center is  re-computed 
as the average of all spectra in the cluster. This
procedure is iterated with the new cluster centers, 
and it quits when no spectrum is re-classified 
in two consecutive steps. The algorithm is simple and fast, 
but it yields different clusters with each random 
initialization
-- the final cluster centers keep some memory
of the original randomly chosen cluster centers. 
This drawback does not interfere with
our purpose of selecting sub-sets of similar spectra 
suitable for averaging because, independently of the
initialization, the clusters always contain
similar spectra.
The algorithm forces all spectra in a class
to be similar to the cluster center, and therefore,
similar among them.
The number of clusters $k$ is arbitrarily chosen but,
in practice, the results are  insensitive to such selection
since only a few clusters possess 
a significant number of
members, so that the rest are discarded.
Figure~\ref{QBCDnumbers} shows the number of elements 
in each class of \qbcd\ spectra resulting from applying
the procedure.
The classes have been sorted and labelled according
to the number of elements, with Class~0 the
most numerous, Class~1 the second most numerous, 
and so on
(percentages are given in Table~\ref{table1}).
Figure~\ref{QBCDclasses} shows the average spectrum corresponding
to the first nine most numerous classes.
The 
classification was carried out using four 
spectral bandpasses containing emission lines 
(the bandpasses are indicated as dotted lines
in Fig~\ref{QBCDclasses}, Class~0, and 
also in Fig.~\ref{classrms}). 
The use of  these particular bandpasses 
emphasizes the contribution of the emission lines
for classification which, otherwise, would be
completely overridden by the continuum.
We select bandpasses throughout the 
full spectral range
to assure that the global trend of the 
continuum is considered when classifying.
There seems to be continuous variation
of properties which, in the end, 
give rise to
a large variety of shapes . From spectra 
without significant emission lines (Class~3),
to spectra with red continuum and emission lines
(Class~7), to blue continua with moderate emission
lines (Class~4). The most numerous Class~0 has  
blue continuum and presents emission lines. 

\begin{deluxetable*}{cccccccc} 
\tablecaption{Properties of \qbcd\ classes and \bcd\ 
classes.\label{table1}}
\tablehead{
\colhead{Galaxy} &\colhead{Class}&\colhead{Fraction\tablenotemark{a}}& \colhead{Stellar}& 
	\colhead{Stellar Age\tablenotemark{c}} &\colhead{Nebular}& \colhead{H$\alpha$ EW\tablenotemark{e}} 
	& \colhead{[N{\sc ii}] EW\tablenotemark{f}}\\
\colhead{Type}& &\colhead{[\%]} &\colhead{Metallicity\tablenotemark{b}}&\colhead{[Gyr]}&
	\colhead{Metallicity\tablenotemark{d}} & \colhead{[\AA]} & \colhead{[\AA]}
}
\startdata
\vspace*{1mm}

\qbcd& 0& 36.1&-0.44$\pm0.03$& 1.7$\pm 0.2$&-0.12$\pm0.18$& 22.3&  5.2 \\ \vspace*{1mm}
& 1& 14.3&-0.39$\pm0.02$& 5.2$\pm 1.4$& 0.05$\pm0.18$&  5.3&  2.5 \\ \vspace*{1mm}
& 2&  9.5&-0.32$\pm0.12$& 1.1$\pm 0.6$&-0.25$\pm0.18$& 50.4&  6.9 \\ \vspace*{1mm}
& 3&  8.5&-0.33$\pm0.02$&11.1$\pm 1.7$&\nodata&\nodata&\nodata\\ \vspace*{1mm}
& 4&  5.3&-0.36$\pm0.08$& 1.0$\pm 0.1$&-0.26$\pm0.18$& 14.5&  1.9 \\ \vspace*{1mm}
& 5&  3.5&-0.68$\pm0.28$& 1.1$\pm 3.0$&-0.40$\pm0.18$& 99.9&  7.5 \\ \vspace*{1mm}
& 6&  3.4&-0.44$\pm0.11$& 1.1$\pm 0.2$&-0.25$\pm0.18$& 31.6&  4.4 \\ \vspace*{1mm}
& 7&  3.1&-0.42$\pm0.04$& 4.1$\pm 1.2$& 0.05$\pm0.18$&  8.4&  3.9 \\ \vspace*{1mm}
& 8&  2.5&-0.37$\pm0.07$& 2.7$\pm 1.7$&-0.01$\pm0.18$& 27.6& 10.0 \\ \vspace*{1mm}
& 9&  2.0&-0.30$\pm0.02$&17.8$\pm 0.4$&\nodata&\nodata&\nodata\\ 
\tableline\vspace{-1.5mm}\\
\bcd& 0& 46.9&-0.33$\pm0.08$& 0.9$\pm 0.1$&-0.38$\pm0.18$& 81.2&  6.7 \\ \vspace*{1mm}
& 1& 12.7&-0.34$\pm0.22$& 1.0$\pm 0.8$&-0.41$\pm0.18$&160.2& 11.7 \\ \vspace*{1mm}
& 2&  7.9&-0.37$\pm0.40$& 0.9$\pm 2.5$&-0.49$\pm0.18$&241.6& 12.4 \\ \vspace*{1mm}
& 3&  7.4&-0.39$\pm0.17$& 1.1$\pm 0.8$&-0.37$\pm0.18$&125.9& 10.7 \\ \vspace*{1mm}
& 4&  6.5&-0.41$\pm0.21$& 0.9$\pm 0.6$&-0.51$\pm0.18$&142.0&  6.5 \\ \vspace*{1mm}
& 5&  3.8&-0.34$\pm0.15$& 1.0$\pm 0.4$&-0.35$\pm0.18$& 93.7&  8.8 \\ \vspace*{1mm}
& 6&  3.5&-0.34$\pm0.54$& 1.1$\pm 4.6$&-0.48$\pm0.18$&333.0& 17.1 
\enddata
\tablecomments{It includes those classes containing 90\% of the 
	galaxies.}
\tablenotetext{a}{Percentage of galaxies represented by the class.}
\tablenotetext{b}{In logarithm scale, referred to the solar metallicity.
	Errors from Monte Carlo analysis in  \S~\ref{metalic}.}
\tablenotetext{c}{Errors from the Monte Carlo analysis in  \S~\ref{metalic}.}
\tablenotetext{d}{In logarithm scale, referred to the solar metallicity. 
	Its error has been taken from  \citet{pet04}.} 
\tablenotetext{e}{Equivalent width of H$\alpha$. No data implies
	line in absorption.}
\tablenotetext{f}{Equivalent width of [N{\sc ii}]~$\lambda$6583. No data implies
	line in absorption.}
\end{deluxetable*}

The scatter among the spectra belonging to a class
depends on wavelength,
and it is largest in the intense emission lines.
As it is illustrated in Fig.~\ref{classrms} with
Class~0 spectrum, it is of the order of 10\% in the 
spectral ranges with absorption lines, and it can be 
of the order of 50\% in the regions having strong emission
lines (see the dashed line in Fig.~\ref{classrms}, which
corresponds to the {\em rms} fluctuations among all the 
spectra in the class divided by the mean spectrum). 

\begin{figure*}
\includegraphics[width=0.7\textwidth,angle=90]{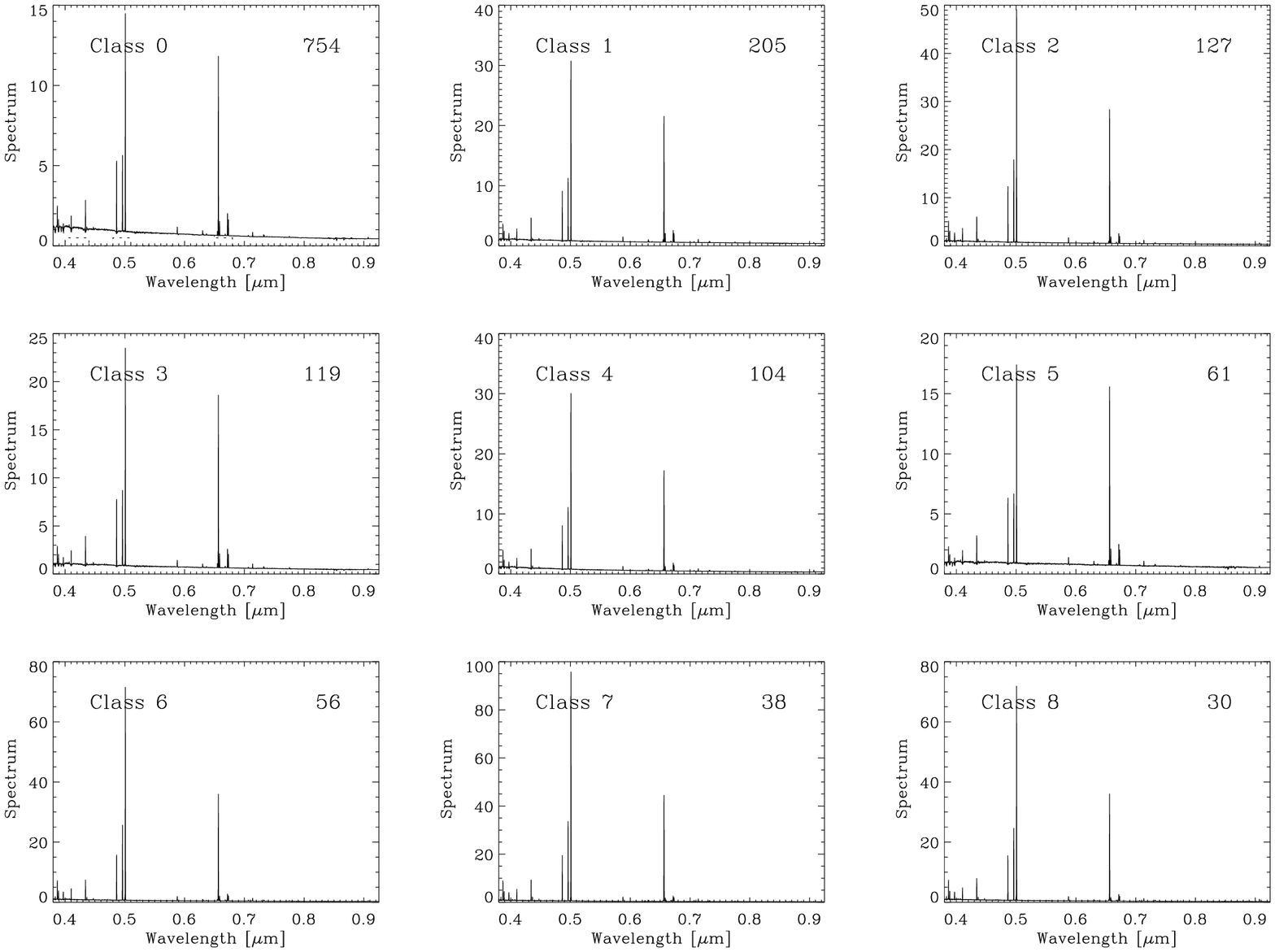}
\caption{Same as Fig.~\ref{QBCDclasses} but for
\bcd\ galaxies.
}
\label{BCDclasses}
\end{figure*}

The same classification procedure  was 
also applied to the control set of \bcd s. 
%
Representative spectra of the most numerous classes
are shown in Fig.~\ref{BCDclasses},
also sorted and labelled according to the number of
galaxies in the class. The most conspicuous 
differences with respect to \qbcd\ spectra are the 
strength of the emission lines, and 
the (barely visible but always present) blue continua.
The fraction of galaxies in each class is listed in Table~\ref{table1}.

\subsection{Green valley \qbcd s}\label{green_valley}
\begin{figure*}
\includegraphics[width=0.7\textwidth,angle=90]{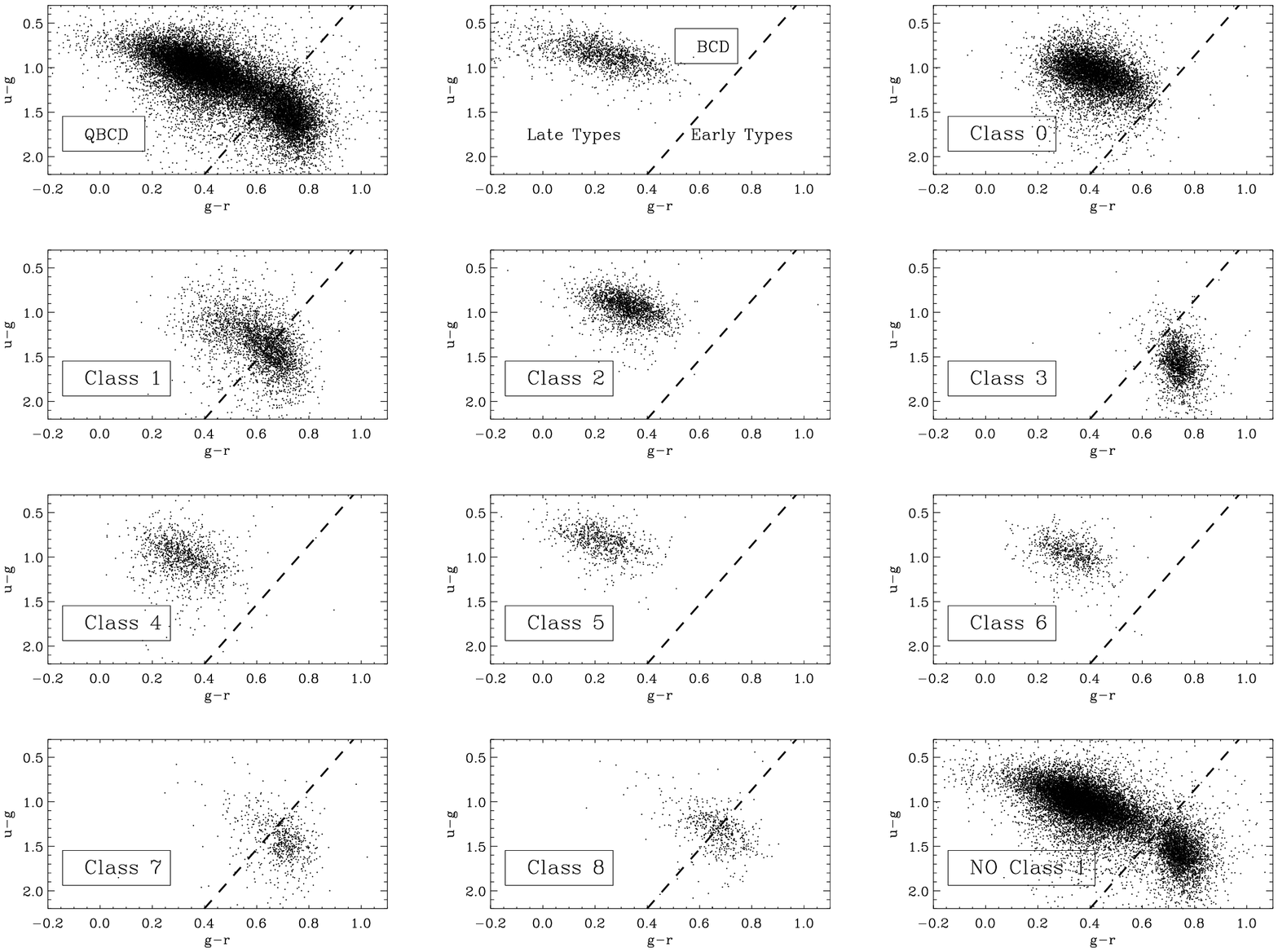}
\caption{
Color vs color scatter plots for the  \qbcd s, the \bcd s,
and the most usual \qbcd\ classes separately. The insets
specify each galaxy group.
The dashed line is the same in all plots, 
and it was worked out in \paperi\
to separate
the blue and the red sequences.
Note that most \qbcd s belong to the blue sequence.
Almost all galaxies in the red clump
correspond to Class~3. Moreover, Class~1 seems
to gather most galaxies in the so-called
{\em green valley} 
between the red and the blue sequences
(c.f. the plots in top 
left and the bottom right corners, which are identical
except that Class~1 has been removed from the latter).
}
\label{color_classes}

\end{figure*}

One of the findings of the SDSS is the existence of a 
galaxy color sequence with well defined
bi-modality \citep[e.g., ][]{bal04}.
\paperi\ shows how 
\qbcd s occupy all the color sequence
between the blue and the red clumps. 
It turns out that the classification has been able to
separate galaxies in the red sequence, 
galaxies in the blue sequence, as well as those in between 
\citep[often refereed to as 
{\em green valley} galaxies, e.g.,][]{sal07}.
Figure~\ref{color_classes}
shows color vs color scatter plots for the \qbcd s, the \bcd s,
as well as the most usual \qbcd\ classes separately.
It also includes the somewhat arbitrary boundary between
the red and the blue sequences worked out in \paperi\
(the dashed line).
The classification does a fair job in splitting 
the galaxies in colors.
Among the most numerous classes, it turns out that
only Class~3 belong to the red sequence. (Class 3
has no emission lines; see Fig.~\ref{QBCDclasses}.) 
In addition, we 
noticed
that Class~1 seems to include all the
{\em green valley} galaxies, i.e., the transition 
galaxies, central to understand how and why galaxies 
move back and forth in the color sequence
\citep[e.g.,][]{spr05,cat06}.
The goodness of this green valley galaxy
selection
method can be appreciated by comparing 
the top left plot with the bottom right plot
in Fig.~\ref{color_classes}. The two of them
include the same \qbcd\ galaxies except for
Class~1. 
A clear gap splits up the red and
the blue clumps.

\begin{figure}
\includegraphics[width=0.49\textwidth]{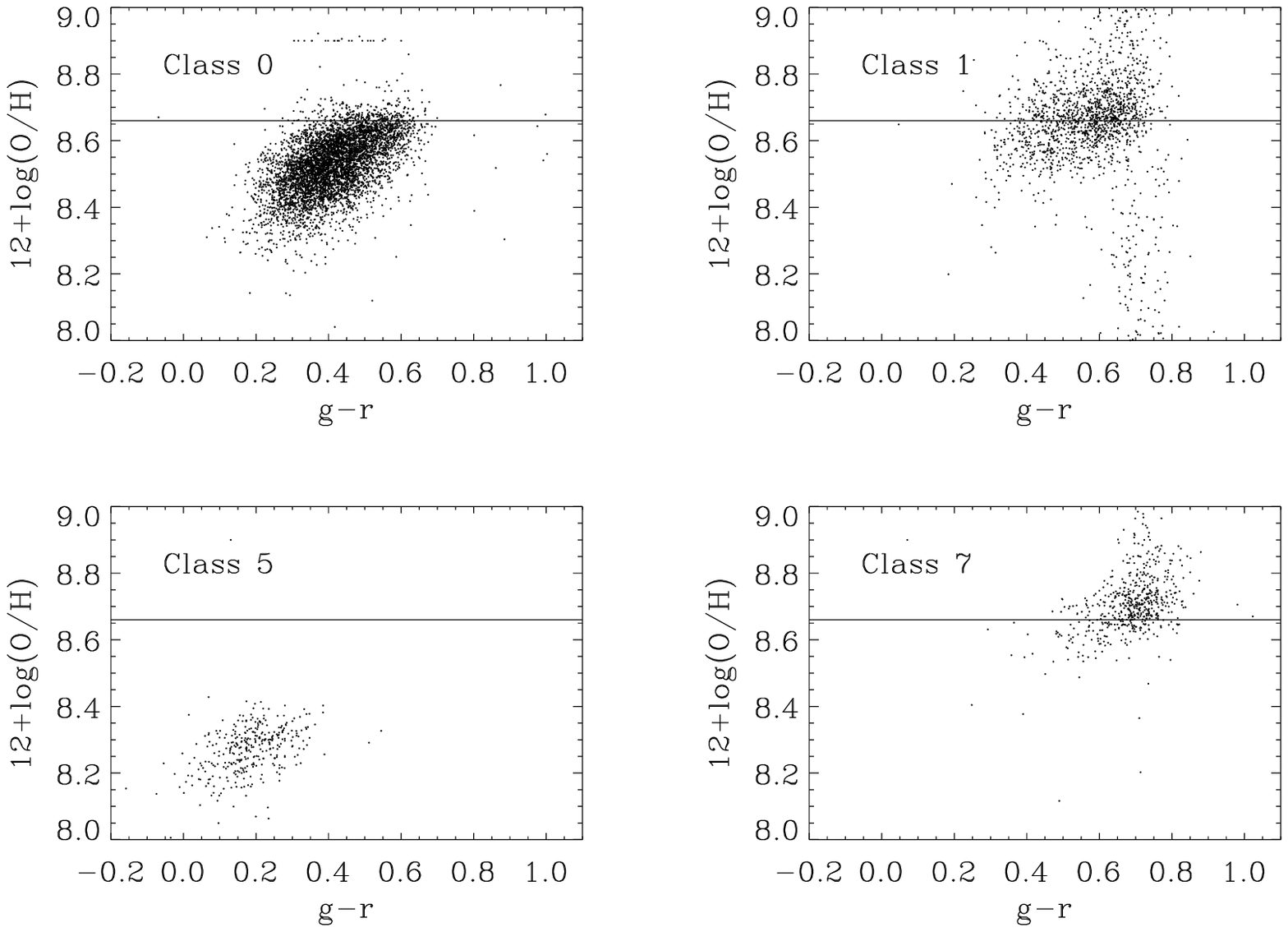}
\caption{
Scatter plot of 
oxygen abundance
vs $g-r$ color for some representative
\qbcd\ classes.
The horizontal solid line corresponds to the solar
metallicity as given by \citet{gre07}. 
Class~3 is not included because it lacks 
of the emission lines needed to compute 
nebular abundances.
}
\label{metal_classes}
\end{figure}

Since different classes have different colors,
and the \qbcd\ present a clear color-(nebular)metallicity 
relationship (\paperi), different classes have different 
metallicities too.  
Scatter plots of metallicity vs absolute magnitude 
are shown in Fig.~\ref{metal_classes}. Here
and throughout we use
the recipe in \citet{pet04} to compute
nebular metallicities from the N2
strong-line ratio (see also \S~\ref{gas_metal}). It upgrades
of the classical calibration by \citet[][]{den02}
used in \paperi\
\citep[][and references therein]{shi05}. 
Late types (e.g., Class~0) are metal poor as compared
to the transition objects included in Class~1. 
These green valley galaxies have solar metallicity. 
The few galaxies in the red clump with emission lines 
(Class~7; see Fig.~\ref{QBCDclasses})
seems to have slightly super-solar 
metallicity~(Fig.~\ref{metal_classes}). 

The k-mean clustering 
classification provides an automatic method to 
identify spectra of green valley galaxies.
It works for \qbcd s, however, there is no clear 
reason why it should be restricted
to them. It may be valid 
for any type of galaxy,
even outside the particular set of dwarfs we
are dealing with.
We elaborate on this possibility
in \S~\ref{conclusions}.
%

%
\section{Determination of stellar metallicities and ages}\label{metalic}

The stellar content of a galaxy can be studied through 
modeling and interpretation of the absorption features
in the integrated spectrum.
The analysis of line-strength indices, mainly those of the Lick system 
\citep{wor94,wor97}, has been the most common approach for 
studying the stellar metallicities and ages.
Most studies have been
focused on early-type galaxies, with stellar populations typically
older than 1\,Gyr \citep[e.g., ][and references therein]{tra98}. The use of
age-sensitive Balmer line indices, e.g., H$\beta$, and metallicity-sensitive
indices, e.g., Mg$b$, allows us to lift
in part the age-metallicity degeneracy affecting the stellar populations of
early-type galaxies \citep[e.g.,][]{wor94,vaz99b}. However, these
line strengths, and particularly the Balmer indices, are not optimal for
analyzing our galaxy spectra, since they are filled-in with nebular emission
(Fig.~\ref{QBCDclasses}). Here we take a different approach,
and ages and metallicities are estimated by direct
comparison of the observed spectra with model spectra 
from stellar population syntheses.
The full spectral range is used simultaneously.
This alternative strategy
has a respectable tradition 
\citep[see ][and references therein]{kol08},
and it allows us to easily overcome the problem 
of emission lines by masking them out. 
Emission lines represent only a small fraction of the spectral 
range, and the rest of the spectrum can be used to 
extract the required information (see below). 
We use an updated version of the models by
\citet{vaz99a}, which provide spectral energy distributions 
of single-age, single-metallicity
stellar populations (SSPs) on the basis of the stellar spectral library MILES
\citep{san06,cen07}.
%
The \miles\ spectra combine, according to 
a Salpeter initial mass function distribution, a suite of
stellar spectra from 0.09\,$M_\odot$ to 100\,$M_\odot$.
They have a resolution of
$\simeq 2.3\,$\AA , a spectral range  
from  3540\,\AA\ to 7410\,\AA , and 
a dispersion of 0.9\,\AA\,pix$^{-1}$. 
\miles\ extends 
the range of ages of \citet{vaz99a}, 
and now it covers  from 0.1\,Gyr to 17.8\,Gyr.
\miles\ spectra
span a range of metallicities\footnote{In the usual
logarithmic scale refereed to the solar metallicity $Z_\odot$, 
i.e., $\log(Z_s/Z_\odot)$ with $Z_s$ the fraction of mass
in metals.}
between $-1.7$ and $+0.2$. The  grid includes
276 SSP spectra, with 46 samples
equispaced in logarithmic time, and 6 steps  in
metallicity. The range of metallicities and ages
fits in well the values to be expected for \qbcd\ 
(see \S~\ref{introduction}). 
As far as the wavelength sampling and wavelength
coverage are  concerned, \miles\ spectral resolution 
is comparable to SDSS (although better), but it misses the reddest
$1800~$\AA\ of the SDSS spectral range.
The uncovered
20\% of the SDSS spectral range is in the near IR, where the
number of spectral lines decreases significantly. 
Keeping in mind all these variables, \miles\ 
meets very well our needs.

\begin{figure*}
\includegraphics[width=0.7\textwidth,angle=90]{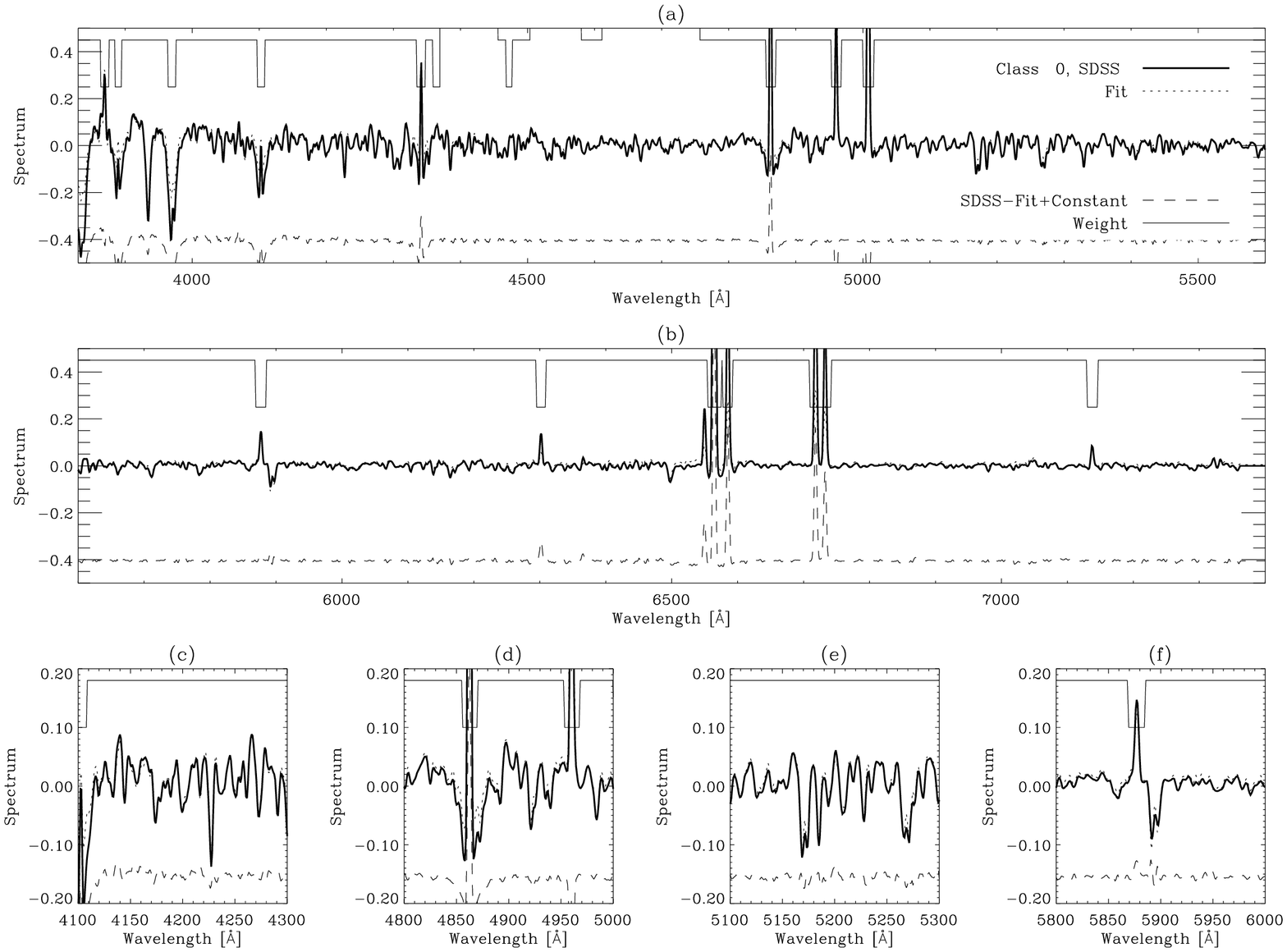}
\caption{
	Observed average spectrum of \qbcd\ 
	Class~0 (the thick solid line) and best fitting 
	\miles\ synthetic spectrum (the dotted line). The
	dashed line corresponds to the residuals
	vertically shifted by an arbitrary amount.
	Panels (a) and (b) show the full spectral range, 
	whereas (c), (d), (e) and (f) zoom into details 
	to appreciate the goodness of the fit.
	The weights of the fits are represented in 
	its own scale as a thin solid line, 
	with the minima corresponding to no contribution, i.e., to
	weight equals zero. 
	The weight goes out of scale in (a) indicating the wavelengths 
	of the three Lick metallic indexes overweighted 
	during fitting ({\tt Fe4383}, {\tt Fe4531}, and {\tt Fe4668}).
	The continuum has been subtracted from both the 
	observed, and the synthetic spectra.
	}
\label{fits}
\end{figure*}

\begin{figure*}
\includegraphics[width=0.7\textwidth,angle=90]{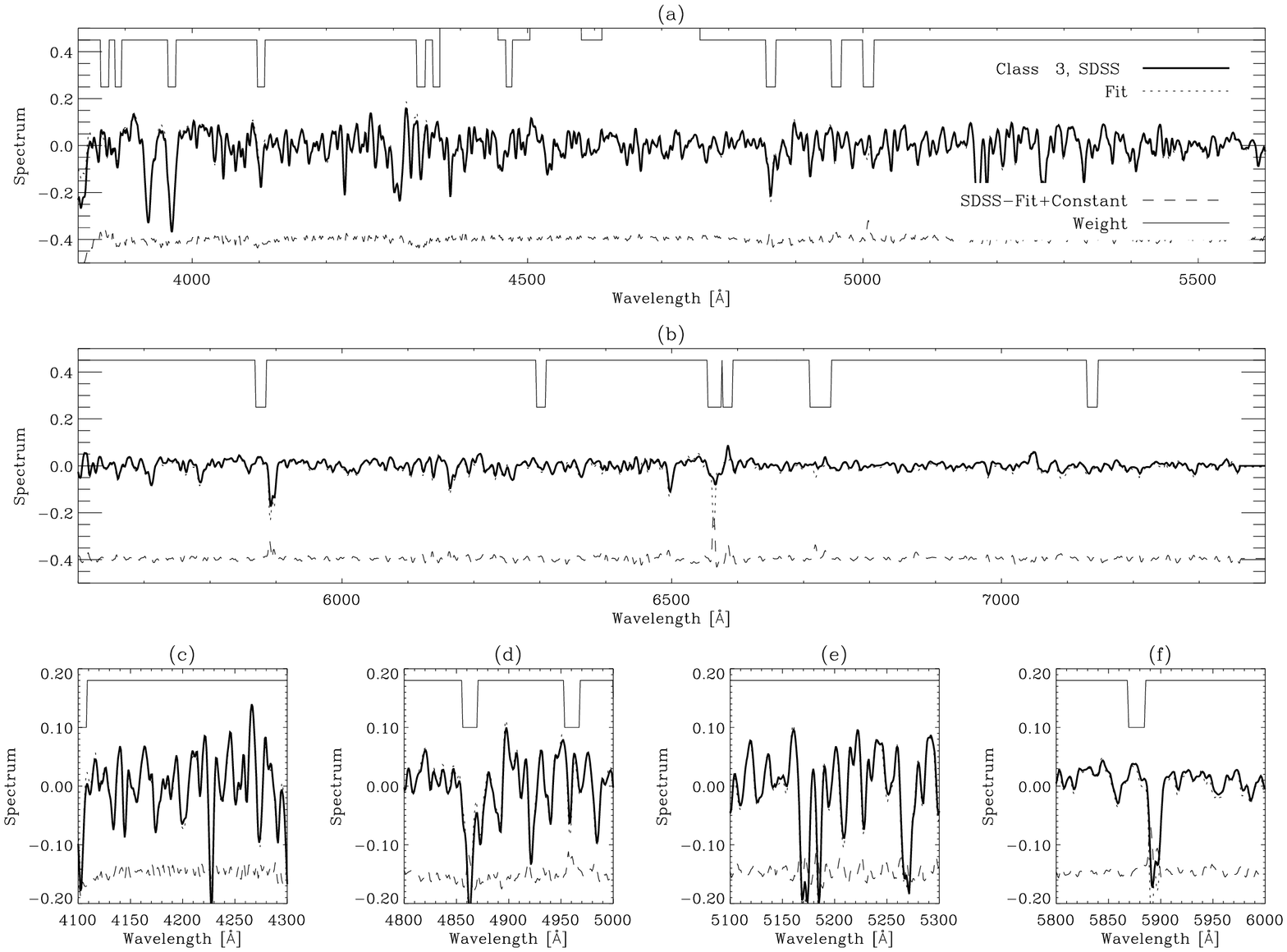}
\caption{
	Same as Fig.~\ref{fits} except that the spectrum corresponds
	to Class~3 \qbcd , 
	i.e., the class corresponding to the red sequence
	galaxies without significant emission lines.
}
\label{fitsb}
\end{figure*}

The fits are carried out by direct comparison of each
average profile representative of a class
with all spectra in the \miles\ 
library, smeared to three
spectral resolutions 
(the original one, the original one plus 2.5~\AA ,
and the original one plus 3.5~\AA).
Considering various broadenings is required 
to account for stellar motions, as well as for the 
difference of spectral resolution between 
MILES SSP and SDSS.
The observed spectrum is compared with each
synthetic spectrum, and closest one in a
lest-squares sense is chosen as best fit.
The comparison was carried out with a 
few constraints which try to minimize potential biases.
(1)~A 100\,\AA\ running-box mean of the original spectra
was removed from observed and synthetic 
spectra. By removing the continua, the results 
of the fits are not very sensitive to the 
extinction, a miscalibration of 
the spectra, the uncorrected differential refraction  
\citep{izo06},
and so on, 
which affect the continua but not 
so much the relative intensity of adjaccent
 spectral lines.
Moreover, it guarantees that ages and
metallicities are inferred from spectral lines,
with negligible contribution from 
the global shape of the continuum.
(2) We assume the observed spectra $O_i$
to be a linear combination of a starburst
spectrum $N_i$ plus an stellar spectrum 
$S_i$,
\begin{equation}
O_i=\alpha N_i+\beta S_i,
	\label{first_def}
\end{equation}
with the underscript $i$ representing the 
$i-th$ wavelength,
and $\alpha$ and $\beta$ being two scaling constants.
The starburst 
spectrum has strong emission lines and little
continuum, therefore, one could simply neglect
the core of the emission lines
when carrying out the fits
(i.e., one could mask out the
emission lines and set $N_i=0$ 
in equation~[\ref{first_def}]).
Here we go a step further so that the 
(small) contamination 
by $N_i$ outside emission lines
is estimated and subtracted out. 
The decontamination procedure works as follows.
At the emission line cores of $O_i$,
the spectrum is dominated by $N_i$ so that,
\begin{equation}
O_i\ (1-w_i)=(1-w_i)(\alpha N_i+\beta S_i)\simeq (1-w_i)\,\alpha\, N_i,
	\label{second_def}
\end{equation}
with $w_i$ a properly chosen weight which is zero
in the emission cores and one elsewhere,
i.e., 
\begin{equation}
w_i=\cases{0& emission lines,\cr
	1 & elsewhere.}
\end{equation}
Using the different classes
of \bcd\ spectra as proxies for $N_i$, we choose
for each $O_i$ the $N_i$ \bcd\ spectrum
that minimizes the appropriate merit function,
\begin{equation}
\chi^2=\sum_i\big[\big(O_i-\alpha N_i\big)^2\,(1-w_i)^2\big],
\label{merit_lines}
\end{equation}
with
\begin{equation}
\alpha=\big[\sum_i\,O_i\,N_i\,(1-w_i)^2\big]\big/\big[\sum_jN_j^2\,(1-w_j)^2\big],
\end{equation}
the latter being just a least squares estimate of the 
scaling factor that best fit the emission lines of $O_i$
once $N_i$ is given. Note that the weight $(1-w_i)$ 
in equation~(\ref{merit_lines}) assures that only the
emission lines contribute to $\chi^2$.
The $N_i$ and $\alpha$ thus derived allows us to compute
the observed spectrum corrected for emission, $O^*_i$, 
\begin{equation}
O_i^*=O_i-\alpha N_i=\beta S_i.
\label{correction}
\end{equation}
The best fitting \miles\ spectrum is obtained
by repeating the same procedure with $O^*_i$
but masking out the emission lines,
i.e., defining the merit function
for each \miles\ spectrum $S_i(t,Z_s)$ as,
\begin{equation}
\chi^2(t,Z_s)=\sum_i\big[\big(O^*_i-\beta S_i(t,Z_s)\big)^2\,w_i^2\big],
	\label{def_chi2}
\end{equation}
where 
\begin{equation}
\beta=\big[\sum_i\,O^*_i\,S_i(t,Z_s)\,w_i^2\big]\big/\big[\sum_jS_j(t,Z_s)^2\,w_j^2\big].
\end{equation}
The expressions explicitly include the dependence
of the synthetic spectrum on the age of the starburst, $t$, 
and the stellar metallicity, $Z_s$, i.e.,   $S_i(t,Z_s)$.
The weight $w_i$ in equation~(\ref{def_chi2}) cancels out the
contribution of the emission lines, rendering the 
correction~(\ref{correction}) of secondary importance.  
The weights $w_i$ are assigned so as to cover the emission line
cores observed in \bcd\ spectra. Examples of these weights are
the (thin) solid lines in Figs.~\ref{fits} and \ref{fitsb}. The 
positions of the minima of these broken lines mark
the wavelengths discarded from the fits.
The fitting procedure described above was also
applied to \bcd\ spectra.
In this case 
we cannot correct for the starburst since \bcd s are used
as template starburst spectra. We just mask out the emission
lines and force $\alpha=0$ in equation~(\ref{correction}).

\begin{figure}
%
%
\includegraphics[width=0.35\textwidth,angle=90]{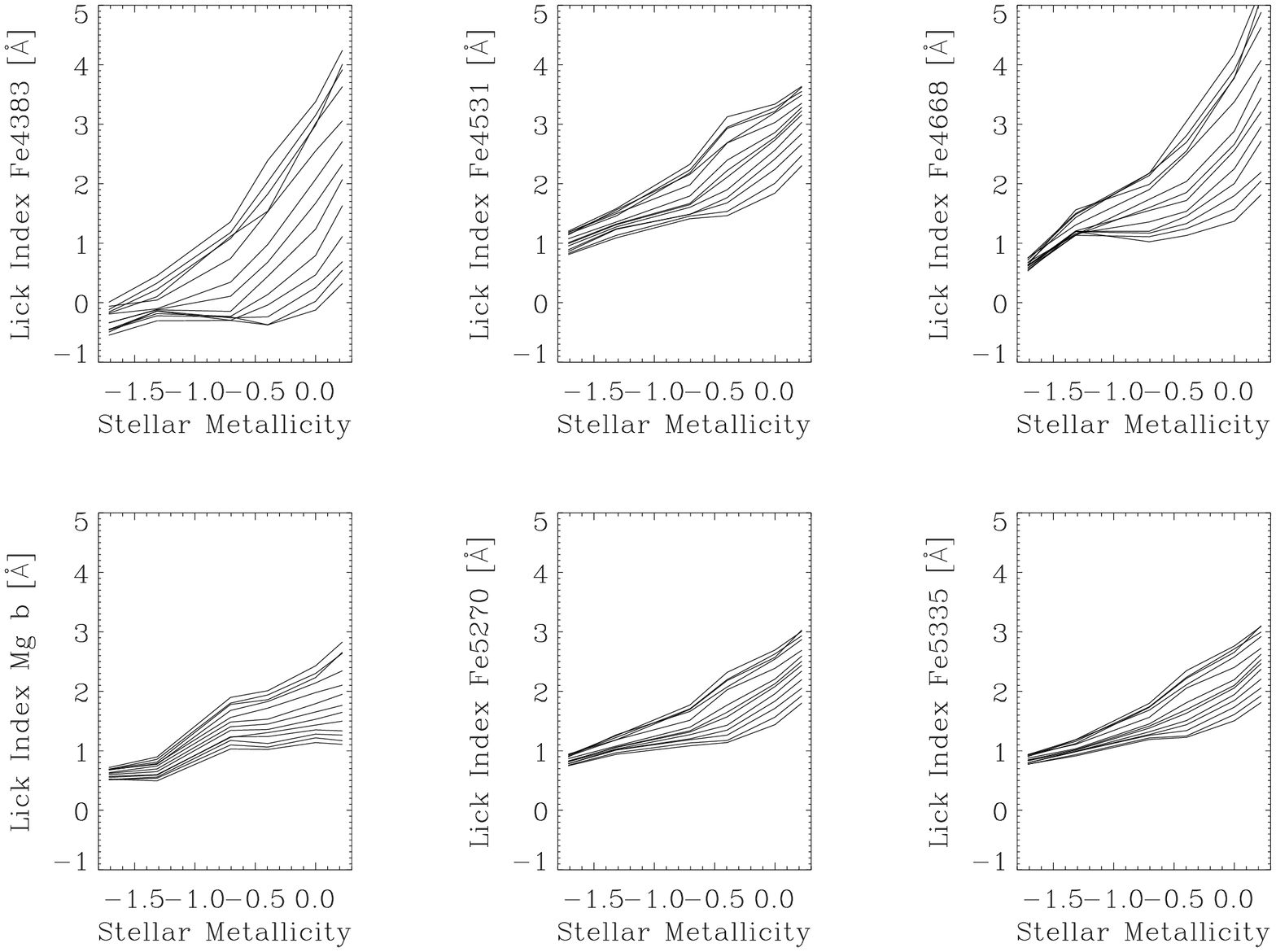}
\caption{Variation with metallicity of  
various Lick indexes in \miles\ spectra. 
Each curve of each plot
corresponds to a constant age.
We only show young populations, 
with ages between 0.5~Gyr 
(the curves of smallest equivalent widths)
and 2~Gyr (the curves of largest equivalent widths).
The wavelength range corresponding to the indexes in the top
row are overweighted in our fits
to break down the metallicity
degeneracy. Other commonly used indexes 
are discarded because they present
less dependence on metallicity for this range
of ages; see the bottom row. 
The index name can be found in the 
ordinate axis labels. The corresponding
bandpasses are defined in  \citet{wor94}.
Metallicities are
given in a logarithm scale referred
to the solar metallicity.
}
\label{index_selection}
\end{figure}

As judged from visual inspection, the best fitting model
spectrum
reproduces very well the observed spectra; see, e.g., 
Fig.~\ref{fits}.  Error bars cannot be assigned using 
traditional methods based on the Hessian matrix
of $\chi^2$, since we do not have a continuous function 
$\chi^2(t,Z_s)$ to compute the partial derivatives with
respect to age and metallicity. 
We resort 
to Monte~Carlo simulations to assign confidence 
intervals.
Gaussian noise with the standard 
deviation of the residuals is added to the 
best fitting model spectrum. This mock observation 
is analyzed as the real observation to get 
an age and a metallicity which, in general, 
differ from those of the best fitting model spectrum. The 
procedure is repeated 1000 times, which provides a 
range of ages and metallicities consistent with the 
best fitting model spectrum and the residual of the fit. 
Confidence intervals thus assigned reveal
a serious problem of degeneracy in the 
metallicity estimate. The standard
deviation of the most common \qbcd\ Class~0
turns out to be 0.25~dex, which 
allows for any metallicity between 
$-0.20$ and $-0.70$.
%
%
%
%
Such degeneracy in metallicity is the young 
stellar population equivalent of the well known
age-metallicity degeneracy appearing in 
early-type galaxy dating \citep[e.g.,][]{wor94b}.
We managed to
sort out the degeneracy problem
by overweighting the contribution of 
those spectral bandpasses that are known
to be particularly sensitive to metallicity.
Specifically, the merit function in 
the definition~(\ref{def_chi2}) is now,
\begin{equation}
w_i=\cases{0& emission lines,\cr
	W& metallicity sensitive bandpasses,\cr
	1 & elsewhere,}
\end{equation}
with $W > 1$ for overweighting.
The band passes were selected 
from the Lick index system \citep{wor94},
which was specifically designed for estimating ages
and metallicities in the integrated light of stellar 
populations. In order to find out which are
the Lick indexes most sensitive to 
metallicity in our domain of ages, we computed 
the variation of the indexes
with metallicity at a given age. 
Some results for the \miles\
library are shown
in Fig.~\ref{index_selection}. 
Among the 21 indexes defined by \citet{wor94}, we
select the three indexes in the top row because
they show the largest variation with metallicity.
Figure~\ref{index_selection}, bottom row,
also includes three other indexes commonly used in 
metallicity studies of early-type 
galaxies \citep[e.g., they are combined
to form the so-called
{$[$MgFe$]$} index; see ][]{gon93,tho03}.
The range of variation is clearly inferior
to the variation of the indexes
that we select.
The bandpasses of the three selected
indexes, {\tt Fe4383}, {\tt Fe4531}, 
and {\tt Fe4668}, are indicated in Figs~\ref{fits}a
and \ref{fitsb}a as the wavelengths 
where the thin solid line representing the weight $w_i$ 
goes out of scale.
We tried with various overweights 
($W=10, 20, 50$ and $100$), to finally
choose  $W=50$ since the trial fits 
indicate that the inferred  metallicity 
does not depend on the actual weight when 
the weights are large enough. 
The use of these overweights improves the
metallicity estimate to a large extent.
We repeated the Monte Carlo simulation 
described above, and the random errors of 
\qbcd\ Class~0 decrease by almost an order of 
magnitude with respect to the case where  
$W$ was set to one.  Table~\ref{table1} includes the 
standard deviation for the ages and metallicities
of all major \qbcd\ and \bcd\ classes. 
They will be employed as 1$\,\sigma$ errors in the 
discusions  along the paper.
The small value of these random errors has been 
independently corroborated by a bootstrap error 
estimate \citep[e.g.,][]{moo03}. 
A caveat is in order, though. These small errors only
indicate that the best fitting \miles\ spectrum is 
well defined, i.e., among the \miles\ set, 
a few spectra reproduce the observation clearly
better than the rest. Our procedure does not 
account for systematic errors, which may dominate the 
error budget  (e.g., is SSP a good
description of our galaxies?). The magnitude of the 
systematic errors is unknown, and ignored in our 
discussions. 
%
%
%

\subsection{Self-consistency of the continua}\label{continua2}

A running mean average was subtracted from both the 
observed and the model spectra to minimize the 
influence of miscalibrations (\S~\ref{metalic}). 
Then our fits are virtually blind to the 
galaxy continua.  The question arises as whether
the ages and metallicities thus derived are or not 
consistent with the observed galaxy continua.

\begin{figure*}
\includegraphics[width=0.7\textwidth,angle=90]{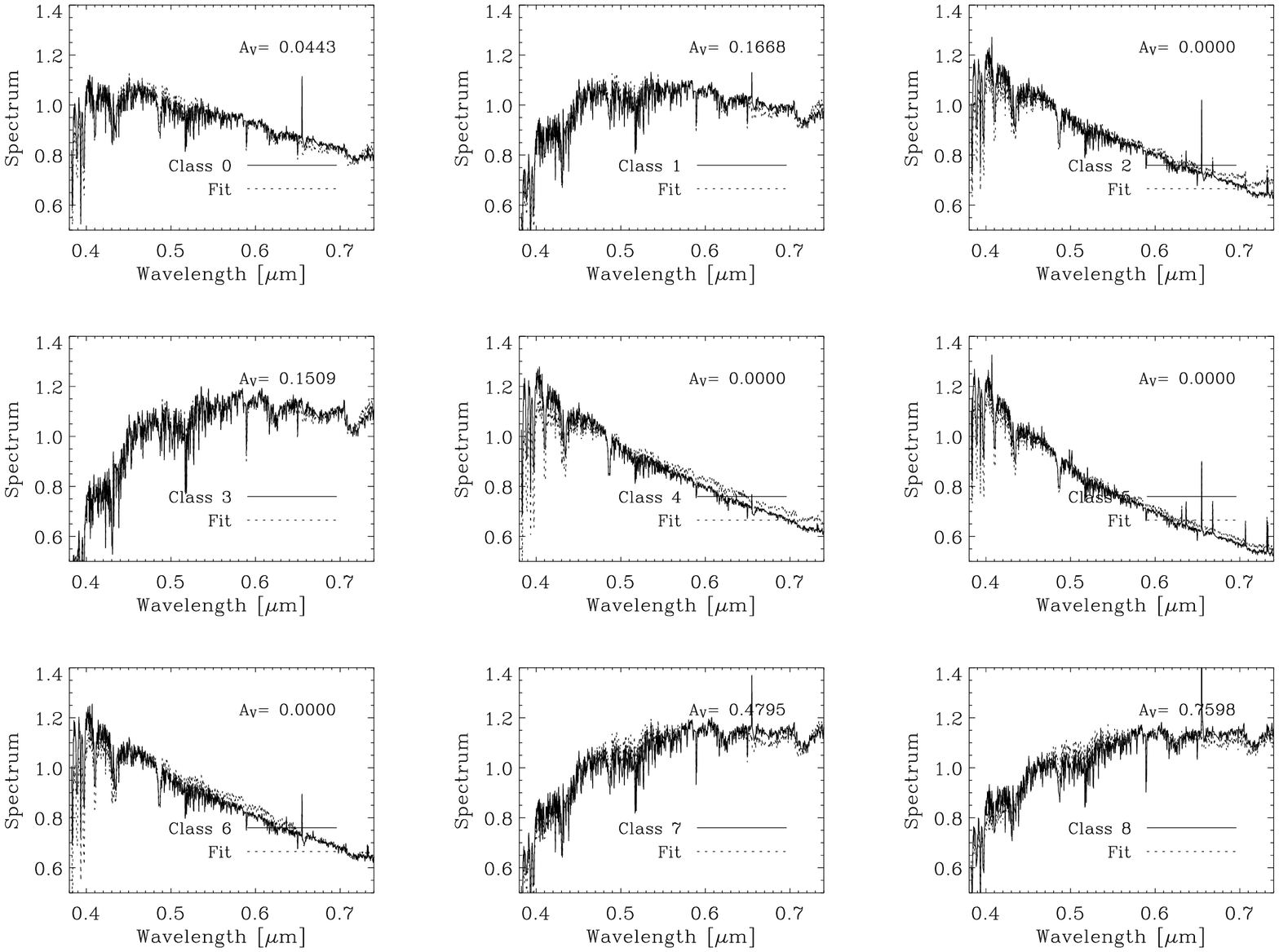}
\caption{Similar to Fig.~\ref{QBCDclasses}, except for
the ordinate scale, magnified to appreciate 
differences between the observed continua (the solid lines),
and the one inferred from fitting absorption lines
(the dotted lines).
The emission lines have been taken out  to
avoid overcrowding of the plots. The insets provide the class number
of the spectrum together with the extinction 
coefficient $A_V$.
}
\label{continua}
\end{figure*}

In order to assign a continuum to the model
spectra, one has to  bring out the continuum 
information removed when subtracting
the running mean average. 
We do it by parameterizing 
the relationship between the actual observed spectrum, 
$o_i$, and  model we fit, $m_i$, 
including the biases that the subtraction of a continuum
removes, i.e.,
\begin{equation}
o_i=m_i\,10^{-(A_i-A_0)/2.5}+\kappa_i,
\label{dust}
\end{equation}
where $A_i$ corresponds to extinction by 
dust\footnote{Both internal, and due to our Galaxy,
since the two of them add up when dealing with
low redshift targets.} defined as usual \citep[e.g.,][]{car89},
and $\kappa_i$ accounts for 
other possible differences not included in the
model.
(Recall that the underscript $i$ parameterizes 
the variation with wavelength.) 
Equation~(\ref{dust}) also assumes that the model
spectrum includes a gray extinction 
given by $A_0$
(incorporated into the global scaling factor $\beta$
that we use for fitting; see equation~[\ref{def_chi2}]). 
As we will show, the
expression~(\ref{dust}) is fully consistent 
with equation~(\ref{first_def}) if one
removes a running mean average
of the observed spectrum, 
$\langle o_i\rangle$.  The spectrum
we fit ($O_i$ in equation~[\ref{first_def}]) is
\begin{equation}
O_i=o_i-\langle o_i\rangle\simeq
(\alpha N_i+\beta S_i)\,10^{-(A_i-A_0)/2.5},
	\label{dust2}
\end{equation}
with the model galaxy spectrum given by 
\begin{equation}
\alpha N_i+\beta S_i=m_i - \langle m_i\rangle.
\end{equation}
We have employed equation~(\ref{dust}) 
assuming that $A_i$ and $\kappa_i$ 
do not vary within the kernel that defines 
the running mean (i.e., $\langle A_i\rangle=A_i$, 
and $\langle \kappa_i\rangle=\kappa_i$).
Neglecting in equation~(\ref{dust2}) 
terms  of the order of, 
\begin{equation}
(\alpha N_i+\beta S_i)\,(A_i-A_0),
\end{equation}
one ends up with,
\begin{equation}
O_i\simeq \alpha N_i+\beta S_i,
\label{nodust}
\end{equation}
which is the approximation used for fitting (equation~[\ref{first_def}]). 
Within this approximation,  
one can re-write equation~(\ref{dust}) as,
\begin{equation}
o_i\simeq \alpha  N_i+\beta S_i + \langle o_i\rangle,
\label{lsfit}
\end{equation}
where 
\begin{equation}
\langle o_i\rangle =\big[1-(A_i-A_0)/(2.5\log{\rm e})\big]\,
\langle m_i\rangle +
\kappa_i,
\label{tbadded}
\end{equation}
is the term to be added to the best
fitting synthetic spectrum, $\alpha  N_i+\beta S_i $, 
to recover the observed spectrum with 
continuum, $o_i$.  
Equation~(\ref{tbadded})
allows us to estimate both $A_i$ and $\kappa_i$ and,
consequently, to complete 
the best fitting model with its continuum.
For lack of better assumption,
we regard $\kappa_i$ as independent of wavelength. 
In addition, the wavelength dependence of $A_i$ is assumed 
to be
known except for a scaling factor, 
parameterized as the extinction in the Johnsson's $V$ band
$A_V$. The ratio $A_i/A_V$ is assumed 
to follow the milky-way law by 
\citet{car89}, modified according
to \citet{mis99} to represent 
the large Magellanic cloud, which
we use as a proxy for low metallicity
extinction law.  (The conclusions below 
remain even if one directly takes the
milky-way extinction law.)
Then  the constants
$\kappa_i$ and $A_V$ can be retrieved from a linear 
least squares fit using equations~(\ref{lsfit}) 
and (\ref{tbadded})
since $o_i$, $m_i$, $\langle m_i \rangle$
and $A_i/A_V$ are all known,
and one can regard $A_0$ as the (wavelength) average extinction.
The comparison between observations and 
model spectra including continuum for 
the nine first \qbcd\ classes is shown in 
Fig.~\ref{continua}.
(The emission lines have been artificially
taken out to better appreciate
differences between observed and model continua.)  
$A_V$  is forced to be non-negative, so that if a
(small) negative number is found in an unconstrained
fit, it is automatically set to zero.
The agreement is  good, in particular, for the most numerous 
classes. Keep in mind that the fitting procedure 
disregards continua, yet, the observed and model 
continua match quite well. The agreement is 
found for low extinctions,
of only a few tenths of magnitudes, 
$A_V=0.18\pm 0.27$.
Moreover, the most populated classes are in the
low extinction range of such interval, e.g.,  
$A_V=0.04$ for Class~0 (see the labels in 
Fig.~\ref{continua}).

\subsection{Why we do not use spectra of individual
	galaxies to estimate ages and metallicities}\label{individual}
Only average spectra are used in our analysis. 
Insufficient signal-to-noise ratio  refrain us from assigning 
ages and metallicities to individual galaxies. The reason stands out 
clearly from the error budget analysis in \S~\ref{metalic}.
The {\em rms} fluctuations of the residual of the fits
are as small as  1.5\%\ (see Figs.~\ref{fits} and \ref{fitsb}) 
and, even in this case, 
the constraint they provide are quite loose. 
The individual SDSS spectra have S/N $\ga 4$  , and
this sole random error would rise the residual 
of any fit to an {\em rms} $\la 25$\%. 
This residual is some 15 times larger than 
the residuals 
of our fits, and such a large error would make
our analysis completely unreliable.
One spectrum is not sufficient. Putting
the same idea in other words;
if the errors of the 
stacked spectra are similar and independent,
averaging at least some $\sim 15^2\simeq 220$ 
spectra is required to get the kind of residual
represented in Figs.~\ref{fits} and \ref{fitsb}.

\section{Gas metallicities}\label{gas_metal}

The gas (or nebular) metallicities of the
different classes of \qbcd s and \bcd s
are estimated using strong~line empirical 
calibration methods.
Specifically, we primarily use the 
so-called  N2~method as provided by \citet{pet04}. 
We cannot employ the more accurate $T_e$~estimate 
because the [OIII]$\lambda$4363 line
required to compute electron temperatures 
\citep[see, e.g.,][\S~3.1]{izo06}
is much too faint in \qbcd s.
Since using strong~line methods is always
controversial 
\citep[e.g.,][]{sta04,sta08,shi05}, we have studied 
some of the potential biases that may arise.
The success of empirical calibrations resides 
in the agreement between the physical conditions of the 
calibration targets, and  those of the galaxies to be 
analyzed. As we discuss in App.~\ref{appa}, the calibration
by \citet{pet04} holds for a fairly large range of 
conditions, broad enough as to encompass the 
different physical conditions to be expected in
\qbcd s and \bcd s. 
Moreover, the various available strong~line 
calibrations give consistent results when applied to our 
spectra.  If the observed differences between the
metallicities of \bcd s and \qbcd s (\paperi , but also 
\S~\ref{introduction} and the forthcoming paragraphs)
were an artifact of using strong~line calibrations,
different calibrations should provide different biases.
However, they coherently show the \qbcd s to be
more metallic that the \bcd s.
Figure~\ref{mainr} presents estimates based
on N2 and O3N2 as calibrated by \citet[][]{pet04}. 
When applied to our spectra, the two of them agree within 
0.1 dex; compare the squares (N2) and  the asterisks (O3N2) 
in Fig.~\ref{mainr}. We 
have also tried with the S23~index as calibrated
by \citet{dia00}, giving results similar to N2 and
O3N2\footnote{Other methods,
like P23 and P \citep{shi05}, cannot be applied
because some of the required emission lines lie
outside the spectral range of the SDSS spectra.}.
In short, the difference of gas metallicity 
between \qbcd s and \bcd s does not seem
to be a bias caused by using the N2~method.
We do not correct for dust extinction to derive
metallicities. This approximation can be 
readily justified since our main calibration 
N2 uses two spectral lines so close in wavelength
that the correction for extinction is truly negligible
($\simeq 0.0006$~dex for a one magnitude extinction). 
We measure the mean extinction for \qbcd s to 
very small (\S~\ref{continua2}), and
\citet[][]{wu08} show how the metallicity
measured in a few representative BCD starbursts 
is not biased by extinction.

The gas metallicities thus obtained
are absolute -- in the end, they 
are based on photo-ionization modeling which relates
the number of observed photons with the number 
of emitting atoms in the photoionized nebula \citep[e.g.,][]{sta04}.
The stellar metallicity, however, is relative to the
solar metallicity. In order to compare gas and stars,
the gas metallicity must be normalized to the solar value. 
This normalization is delicate
and may bias the comparison, particularly in this moment
when a major revision of the solar metallicity scale has 
occurred \citep{all01,asp05b,gre07}, and it is 
not consistently implemented in modeling. For this
reason, we feel compelled to discuss our use of the
modern oxygen abundance for gas metallicity normalization,
\begin{equation}
12+\log({\rm O/H})_\odot=8.66\pm 0.05,
\label{new_solar_abu}
\end{equation}
despite the fact that the \miles\ spectra used
in our stellar metallicity estimates are based on 
a stellar library whose metallicities date back to
pre-revision days \citep{cen07,leb04,cay01}.
The modification of the of solar metallicity had
to do with improved modelling -- NLTE effects and 
realistic 3D hydrodynamical model atmospheres have 
been incorporated into the analysis \citep{asp05b}. 
Since the observed
solar spectrum has not been modified, the revision
simply re-labelled it with a different metallicity.
For the sake of argumentation, assume that the
 spectrum of one of our galaxies has
solar metallicity. Then it has the metallicity
corresponding to the oxygen abundance in
equation~(\ref{new_solar_abu}), rather than
the metallicity originally assigned to it.
Consequently, the use of \miles\ spectra
to estimate stellar metallicities is consistent
with using equation~(\ref{new_solar_abu})
for the solar metallicity that
normalizes the absolute gas metallicity
inferred from emission lines.
Table~\ref{table1} lists the relative N2
gas metallicities thus computed, together with
the equivalent widths of two lines
used in such estimate (H$\alpha$
and [N{\sc ii}]~$\lambda$6583).

\section{Gas metallicity vs stellar metallicity}\label{metal_vs_metal}

In this section
we compare the 
stellar metallicities worked out in \S~\ref{metalic}
with the gas metallicity derived
from emission lines in  \S~\ref{gas_metal}. 

\begin{figure}
\includegraphics[width=0.49\textwidth]{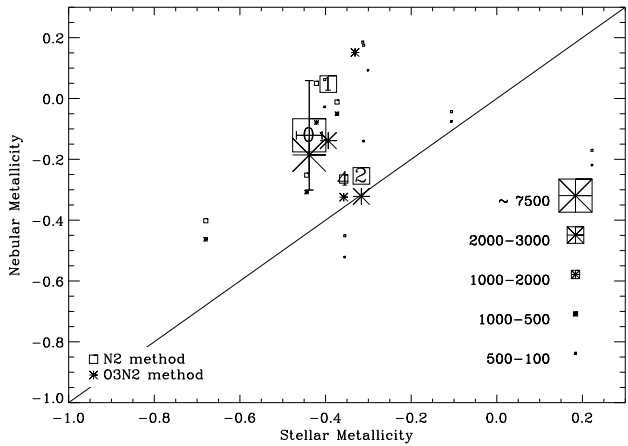}
\caption{Emission line based
	metallicity vs stellar metallicity
	for the set of QBCD classes. The size of the symbol
	indicates the number of galaxies represented by the 
	class, as specified in the inset. Boxes and 
	asterisks correspond to  two different estimates
	of emission line oxygen metallicity. 
	The numbers inside the square symbols 
	identify the major classes.
	Error bars for Class~0 are shown for reference,
	and they are similar to those of all major 
	classes; see Table~\ref{table1}.
	}
\label{mainr}
\end{figure}

\begin{figure}
\includegraphics[width=0.49\textwidth]{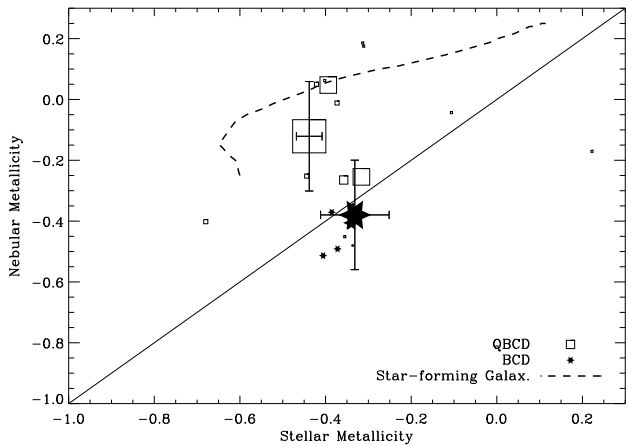}
\caption{Same as Fig.~\ref{mainr} but for both \qbcd s 
(the square symbols) and \bcd s (the star symbols). 
The dashed line shows the relationship 
for star-forming galaxies inferred from the works
of  \citet{tre04} and \citet{gal05} (see main text).
The size of the symbol scales with the number 
of galaxies in the class.
	Our estimate of Class~0  
	error bars is  shown for reference.
}
\label{mainrc}
\end{figure}

The scatter plot gas metallicity vs stellar metallicity
is shown in Fig.~\ref{mainr}. It includes all 
\qbcd\ classes 
except Class~3, which has no emission 
lines. Several features are notable.
First, the stellar metallicities 
of the most representative Classes (0 and 1)
are systematically
smaller than the nebular metallicities\footnote{
Since these metallicities are refereed to the 
solar value, we can directly compare the global
metallicity by mass provided by the spectral
fitting procedure, with the oxygen metallicity 
by number inferred from emission lines.
They correspond to the same
quantity under the implicit assumption
that the relative metal abundance of our
targets follow the solar composition. This
should be a good approximation for
dwarf galaxies \citep[e.g.,][]{mic08}.
}. This
result holds even when the (large) error bars
of our metallicity estimates are taken into account.
Figure~\ref{mainr} includes the error bars assigned 
to the metallicities of Class~0, and
it is clear that the retrieved stellar and
nebular metallicities disagree. The stellar metallicity
error bar has been taken 
from the Monte~Carlo simulations described in 
\S~\ref{metalic}, and is listed in Table~\ref{table1}.
As for the emission line
metallicity, we take 0.18~dex inferred by \citet{pet04}
from the dispersion of the N2 based metallicity
when compared with the more precise $T_e$ method.
(See also App.~\ref{appa}.)
Error bars are similar for other classes.
They have not been included to
avoid cluttering Fig.~\ref{mainr}, but they are listed
in Table~\ref{table1}.
Note that contrarily to the  behavior of Classes~0 and 1, 
Classes~2 and 4 present similar stellar and nebular 
metallicities. This is not so much due to a change of 
stellar metallicity, but to a significant decrease 
of the nebular metallicity.
The same kind of agreement at low metallicity
occurs for all \bcd\ classes.
Figure~\ref{mainrc} includes 
the scatter plot of nebular vs 
stellar metallicity for the \bcd\ classes
(the solid star symbols).
Stellar and nebular metallicities agree
in this case, discarding serious systematic 
errors biasing our conclusion.
To be more
precise, assuming 
that systematic errors in 
\qbcd s and \bcd s metallicities are similar,
the 
\qbcd s and the \bcd s
tend to have the same stellar
metallicity but different nebular metallicities. 
Taking the most numerous Class~0
to represent them
(i.e., the largest symbols with error bars
in Fig.~\ref{mainrc}),
the stellar metallicities $Z_s$ of \qbcd s and 
\bcd s are similar, 
\begin{equation}
\log(Z_s/Z_\odot)_\qbcd\simeq\log(Z_s/Z_\odot)_\bcd, 
	\label{equal_metal}
\end{equation} 
and also similar to the nebular metallicity of \bcd s, 
\begin{equation}
[{\rm O/H}]_\bcd\simeq\log(Z_s/Z_\odot)_\bcd, 
\label{equal_metal_2}
\end{equation} 
which differs from the nebular metallicity of \qbcd s,
\begin{equation}
[{\rm O/H}]_\qbcd
	\simeq[{\rm O/H}]_\bcd+0.35.
\label{diff_metal}
\end{equation}
As usual, we have employed the notation where
$[{\rm O/H}]=\log({\rm O/H})-\log({\rm O/H})_\odot$,
and $Z_\odot$ stands for the solar metallicity.
Equations~(\ref{equal_metal}), (\ref{equal_metal_2}), 
and (\ref{diff_metal}) combined
yield,
\begin{equation}
[{\rm O/H}]_\qbcd\simeq
\log(Z_s/Z_\odot)_\qbcd + 0.35.
\label{diff_metal_2}
\end{equation}
All the above identities
have an uncertainty of the
order of 0.2~dex, which is large but does not
invalidate the trends.

Figure~\ref{mainrc},  the dashed line,  includes
the relationship between nebular
and stellar metallicities corresponding to a large
set of SDSS star-forming galaxies. 
We have inferred such a
relationship by combining the medians
of the mass-metallicity relationships by \citet{tre04}
and \citet{gal05}. \citet{tre04} derive metallicities
from emission lines, whereas \citet{gal05} metallicities
refer to the luminosity weighted stellar metallicity.
The emission line metallicities are referred to the solar
metallicity using the solar oxygen abundance employed
by \citet{cha01}, which is the code behind \citet{tre04} 
estimates. 
Figure~\ref{mainrc} shows how the nebular metallicities 
are  systematically larger than the stellar metallicities for 
galaxies of the same mass, and such difference
is similar to the one we find for  \qbcd s
(equation~[\ref{diff_metal_2}]).
We cite this disagreement as a consistency
test for our metallicity determinations since 
both \citet{tre04} and \citet{gal05} derive
metallicities using tools different and
more elaborated than the ones used here.
If an unknown bias is causing 
the differences between nebular and stellar metallicities
in \qbcd s, it does not seem to be due to our specific 
simplifying  hypotheses.

The error bars employed so far correspond 
to 1$\,\sigma$, or 68\% 
confidence level. If we use 2$\,\sigma$ instead,  
we cannot discard the agreement between 
the gas and the stellar metallicities of \qbcd s
\citep[in the case of N2 based nebular metallicities,
2$\,\sigma\simeq 0.41$~dex;][]{pet04}. 
Note, however, that the error bars used 
for the nebular metallicity are rather 
conservative. We find that the metallicities 
inferred from  O3N2 and N2 agree 
(\S~\ref{gas_metal} and Fig.~\ref{mainr}), and the
scatter of the O3N2 relationship found by 
\citet[][]{pet04} is significantly smaller than
that of N2 method (2$\,\sigma\simeq 0.25$~dex for 03N2). 
Moreover, the scatter found by \citet{pet04}
corresponds to  individual extragalactic
H~{\sc ii} regions. Part of such scatter
have to be of random nature, 
and it cancels when averaging many different 
regions or, as we do, many different galaxies.  
Only the (unknown) systematic part of the error 
would be of relevance in our case. 

\section{Age of the stellar component}\label{ages_sect}
\begin{figure}
\includegraphics[width=0.49\textwidth]{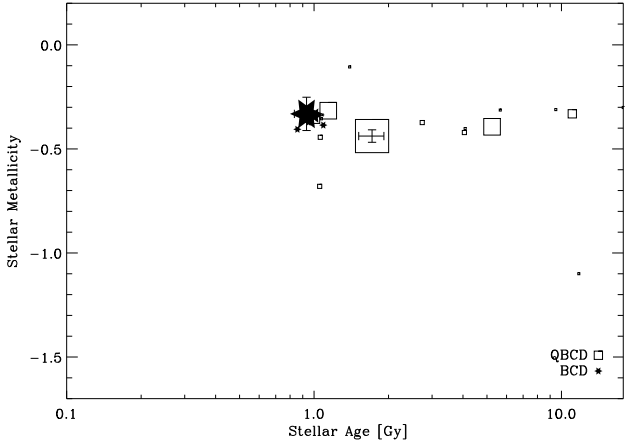}
\caption{Stellar metallicity vs stellar age scatter plot for both \qbcd s 
(the square symbols) and \bcd s (the solid star symbols). 
Error bars for Class 0 \qbcd s and \bcd s are included.
The size of the symbols codes the number of galaxies 
in the class, like in
Fig.~\ref{mainr}.
The range of the axes corresponds to the full range of ages 
and metallicities spanned by the \miles\ library.
Note how the assigned ages and
metallicities occupy a well-defined narrow region
among of the possible solutions.
}
\label{age}
\end{figure}

The absorption line spectrum fitting procedure in 
\S~\ref{metalic} provides ages and metallicities for the 
stellar component of the galaxies.
Figure~\ref{age} shows the metallicity vs age 
scatter plot 
corresponding to the two sets of galaxies, 
\qbcd s and  \bcd s. We find that
\qbcd s are systematically older than \bcd s. 
\qbcd s have ages in excess of 1~Gyr whereas
all \bcd s have ages inferior to (but close to) 1~Gyr.
The case of \qbcd\ Class~3 is worthwhile mentioning 
separately (the oldest age in Fig.~\ref{age}, of the
order of 11~Gyr). 
It corresponds to the \qbcd\ galaxies without 
emission lines (Fig.~\ref{QBCDclasses}), 
which form the red clump
of the color sequence (Fig.~\ref{color_classes}).
These properties hint at Class~3 being early type 
galaxies, and the age we find is also consistent with this
possibility.
Such a very old origin of the stellar population
of Class~3 is very well constrained according to our 
error analysis -- 
see Table~\ref{table1}.

According to the conjecture we are 
examining in the paper (\S~\ref{introduction}), 
\qbcd s undergo successive starbursts that transform then to
\bcd s for short periods \citep[lasting only 10 Myr or so; see,
e.g.,][]{mas99}.
The number density of \bcd s and \qbcd s requires
the bursting phase to appear, statistically, every 0.3 Gyr. 
The fact that this timescale differs from the
age we assign to the \qbcd s is not at odds
with the conjecture. The stellar population
we observe now has been produced
not just during the  last starburst, 
but during several bursts.
It is the luminosity weighted age of these
populations what we have estimated, which
has to exceed the age of the last starburst.
The fact that the age of the stellar 
population of \bcd s is shorter but not very different 
from the age of \qbcd s adds on to this picture. 
If \bcd s and \qbcd s are
basically the same galaxies, but the \bcd s happen
to be in a phase of enhanced star formation activity, 
then the underlaying stellar populations must have 
similar properties.
This turns out to be the case.
The metallicities are similar (\S~\ref{metal_vs_metal} 
and equation~[\ref{equal_metal}]),
and \bcd\ stellar ages are shortened due
to the strength of the current starburst.

The fact that the age of the stellar 
population of \bcd s is much larger than the
age of a typical starburst supports that \bcd\ are not
forming stars for the first time. They have an 
underlaying stellar population much older than 
a starburst.

\section{Metal enrichment and Star Formation}\label{chemical}

This section analyzes the difference between
the observed nebular and stellar metallicities of \qbcd s. 
We will find that the nebular metallicity of \qbcd s 
seems to be much too high to be representative 
of the galaxy as a whole. Should it be 
representative, the \qbcd\ galaxies would 
have been forming stars during the last few Gyr at 
an unobservedly large SFR. As we conjectured
in \paperi , the metallicity of \qbcd s 
inferred from emission lines probably 
represents a small fraction of the galactic 
gas, locally contaminated by recent
starbursts.

In principle,
the difference between stellar metallicity and 
gas metallicity found in \S~\ref{metal_vs_metal} 
may be explained in terms of the chemical enrichment of 
the ISM  during the time span 
between the formation of the stars and the present 
epoch. Assume that the chemical 
enrichment has followed a closed-box evolution. 
(The consequences of an open-box evolution will be 
discussed later on.) In this case the conservation 
of metals imposes the following constraint
between the mass of stars $M_s$, 
the mass of gas $M_g$, the metallicity of the stars
$Z_s$, the metallicity of the gas $Z$, and
the yield $y$,
\begin{equation}
Z\,M_g + Z_s\, M_s = y\, M_s+ Z_0\, M.
\label{chem1}
\end{equation}
(We will use without explicit citation 
well known results from the theory of 
chemical evolution of galaxies; see, e.g.
\citeauthor{tin80}~\citeyear{tin80};
\citeauthor{pag97}~\citeyear{pag97}.)
The left hand side of equation~(\ref{chem1})
gives the amount of metals now existing in stars 
and gas, which is equal  to the metals 
created by stars, $y\, M_s$, plus the metals
existing at the beginning of the 
starburst, $Z_0\, M$, where $M$ stands
for the total mass  in the star-forming 
closed-box, and $Z_0$ represents
the initial metallicity of the gas.
Equation~(\ref{chem1}) can be rewritten in a 
more convenient form,
\begin{equation}
\mu={{y+Z_0-Z_s}\over{y+Z-Z_s}},
\label{chem2}
\end{equation}
with $\mu$ the mass fraction of gas,
\begin{equation}
\mu=M_g/M.
\end{equation}
Also from the theory of closed-box evolution,
\begin{equation}
Z-Z_0=-y\ln\mu ,
\label{chem3}
\end{equation}
and
\begin{equation}
\mu=1-{{(1-R)\,\sfr\,t}\over{M}},
\label{chem4}
\end{equation}
with \sfr\ the average Star Formation Rate 
during the past time interval $t$, 
and $R$ the fraction of stellar mass that returns to 
the  ISM rather than being 
locked into stars and stellar remants.
Equations~(\ref{chem2}) and (\ref{chem3}) combined
provide the difference of metallicity 
between gas and stars,
\begin{equation}
Z-Z_s=-y\,\big[{{\ln\mu}\over{1-\mu}}+1\big].
\label{chem5}
\end{equation}
In principle, the mass of the starburst $M$ is
unknown, but our ignorance can be parameterized 
using  a scaling factor $f$ between $M$ and   
the mass of stars in
the galaxy at present $M_*$, i.e.,
\begin{equation}
M=f\,M_*.
\label{chem6}
\end{equation}
Although it is not a primary observable,
the stellar mass content of a galaxy 
can be inferred from its observed luminosity
and color. 
Using the calibration modeled by \citet{bel01}, 
the color transformations  between Johnsons's colors and 
SDSS colors by \citet{jes05},
and the typical colors of the \qbcd\ 
candidates in \paperi ,
one finds a relationship between the
stellar mass in solar mass units, 
and the absolute magnitude in 
the SDSS $g$~color,
\begin{equation}
\log(M_*/M_\odot)\simeq -0.50\, g+0.35\,.
	\label{sfr1}
\end{equation}  
The parameters $R$ and $y$ are constrained
by the stellar evolution models, so that
$R\simeq 0.2$ \citep[e.g.,][]{tin80,pag97,apa04}
and, for oxygen, $y\simeq 4\times 10^{-3}$ 
\citep[][and references therein]{dal07}.
Then given the age, the \sfr ,  and the mass (i.e., $f$)
of a starburst,
equations~(\ref{chem4}), (\ref{chem5}) and (\ref{chem6}) 
allows us to predict the  difference of metallicity
between gas and stars, which is the parameter measured in 
\S~\ref{metal_vs_metal}. 
Consequently, we can use them to estimate the mass of the 
starburst and/or the \sfr\ required to explain the observed
metal enrichment of the gas.
This is what we 
do next.

\begin{figure}
\includegraphics[width=0.49\textwidth]{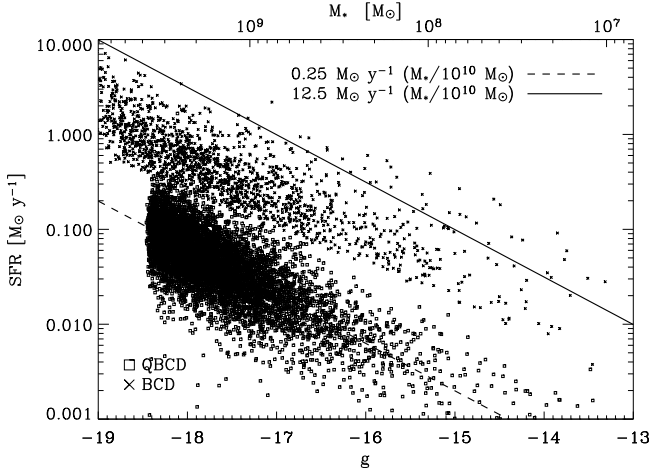}
\caption{Star Formation Rate (SFR) vs SDSS 
$g$~absolute magnitude.
Both \qbcd\ galaxies (Class~0; the square symbols) and
\bcd\ galaxies (full set; the times symbols) are included.
The dashed line represents the average \qbcd\
\sfr , i.e., 
$\sfr\simeq 0.25\cdot (M_*/10^{10} \,M_\odot)$. 
The solid line corresponds to a \sfr\ 50 times
larger.
The stellar mass of the \qbcd\ galaxies derived from
the $g$~magnitude is also included in the scale on top of 
the figure.
 }
\label{sfr_uno}
\end{figure}

One can estimate the present \sfr\ of the \qbcd\ galaxies
from their H$\alpha$ emission. We have used the  
prescription in \citet[][]{ken98}, which gives the SFR 
as a function of the  H$\alpha$ luminosity. 
We derive the luminosity from the observed H$\alpha$ 
equivalent width, and the absolute luminosity of the 
galaxies in the SDSS $r$ bandpass, approximately
centered at the H$\alpha$ wavelength. The transformation
between magnitudes and fluxes has been carried out
keeping in mind that the SDSS color system is an 
AB system \citep{smi02}, which renders,
\begin{equation}
{\rm SFR}\simeq \gamma\,{{W_{{\rm H}\alpha}}\over{100\,{\rm \AA}}} 10^{-0.4(r+19.0)}\,
M_\odot\,{\rm yr^{-1}},
	\label{present_sfr}
\end{equation}
with $W_{{\rm H}\alpha}$ the equivalent width in \AA , 
$r$ the integrated
absolute $r$ magnitude, and $\gamma$ 
the fraction of galactic light contributing to the
the starburst.\footnote{Note that 
the SDSS equivalent width corresponds to 
a spectrum taken at the center of the galaxy (\S~\ref{spectra}). Using
the integrated luminosities to estimate $H\alpha$ 
fluxes assumes that the star-forming burst observed
at the galaxy core extends evenly
throughout the galaxy. 
The factor $\gamma$ accounts for the case where the 
starburst affects only a faction $\gamma$ of the 
galaxy.}
Figure~\ref{sfr_uno} shows the \sfr s  inferred from 
equation~(\ref{present_sfr}) for the 
individual \qbcd\ galaxies in Class~0 
(the square symbols).
We have used $\gamma=0.5$, as a reasonable upper limit for
the extent of the starburst,
but using 0.1 or 1 do not change any of the conclusions
discused below. 
Typically, the \sfr\ is $0.1\,M_\odot\,$y$^{-1}$ 
for the brightest \qbcd s, i.e., when $g\simeq -18.5$.
Figure~\ref{sfr_uno} also includes
the \sfr\ of \bcd s --  in this case
$\gamma=1$ to acknowledge that the starburst is spreadout 
all over the galaxy.

\begin{figure}
\includegraphics[width=0.49\textwidth]{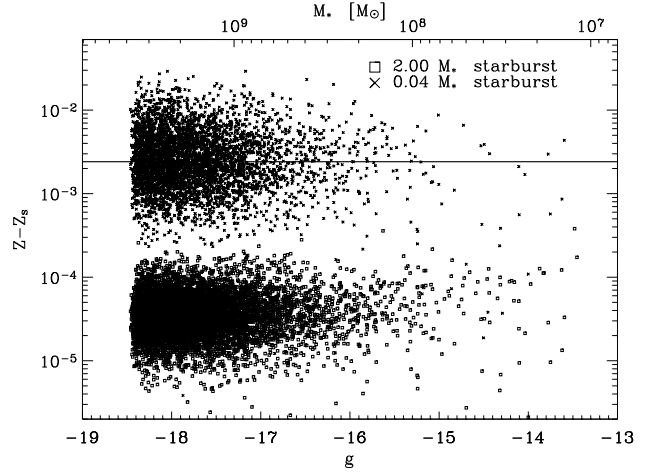}
\caption{Difference between nebular and stellar 
metallicities
($Z-Z_s$) predicted if the chemical evolution of 
Class~0 \qbcd s follows a closed-box model. 
The difference is plotted vs the
absolute $g$-color magnitude (bottom axis) 
and vs the stellar
mass of the galaxy ($M_*$, top axis). 
The horizontal solid line corresponds to the
observed difference, which can be reproduced
only if a small fraction of the galactic 
gas is polluted with metals (see the main text). 
}
\label{deltaz}
\end{figure}

Figure~\ref{deltaz} shows the difference between nebular 
and stellar oxygen metallicities predicted by the closed-box 
evolution of Class~0 \qbcd\ galaxies. 
We assume $t$ to be the age of the stellar population
derived in \S~\ref{ages_sect}.
Equations~(\ref{chem4}), (\ref{chem5}), (\ref{chem6}),
(\ref{sfr1}), and (\ref{present_sfr}) were used with
$f=2$ (the square symbols)  and $f=0.04$ (the times symbols).
The case $f=2$ represents a galaxy-wide starburst
able to pollute with metals 
the whole galactic gas. ($f=2$ assumes
the same amount of mass in gas as the
mass in stars, which is reasonable
for low surface brightness dwarf galaxies like
our \qbcd s. According to \citet{sta92},
they have one $M_\odot$ of H{\sc i} gas 
per solar luminosity, which corresponds to
$f\simeq 3$.) 
In this case the predicted difference of metallicity
is too low to account for the observed difference 
(\S~\ref{metal_vs_metal}), 
which is represented in Figure~\ref{deltaz} as a horizontal
solid line. The amount of metals produced at the
current \sfr\ during the age of the 
starburst is insufficient to effectively 
contaminate the whole ISM of the galaxies. 
If, on the other hand, the same starburst pollutes 
only a small fraction of the galactic gas ($f=0.04$, 
or a factor 50 smaller than the previous case), then
the predicted and the observed metallicities 
agree (Fig.~\ref{deltaz}, the times symbols).
Consequently, the observed $Z-Z_s$
can be explained if 
the gas from which we infer the metallicity
represents only a small fraction of the 
total galactic gas. 

Agreement between observed and model metallicities 
can be also reached if the \qbcd\ galaxies 
had an average \sfr\ during $t$ much larger 
than the present one. The predicted closed-box evolution 
depends on the ratio $\sfr/f$ rather than on
\sfr\ and $f$ separately, and a decrease
of $f$ is equivalent to an increase of \sfr ; 
see equations~(\ref{chem4}) and (\ref{chem6}).
However, the required increase of \sfr\ is too 
high to be sensible. It would have to be 50 times
larger than the observed ones. This level of
{\em continuous} star formation activity 
during the last two Gyr is unreasonably 
high; it is shown as a  solid line in 
Fig.~\ref{sfr_uno} and it corresponds to 
$\sim 5\,M_\odot\,$y$^{-1}$ for the brightest \qbcd\
galaxies. It is larger that the already large \sfr\
observed in our \bcd\  galaxies (see Fig.~\ref{sfr_uno},
the times symbols).

So far our argumentation has considered 
closed-box chemical evolution. If the box
is opened 
both infall of (low metallicity) gas from the 
intergalactic medium, and outflows of
metal rich SNa ejecta are
to be expected \citep[e.g.,][and references
therein]{gar02,dal07}, and both processes
reduce the
metallicity of the gas with respect to the predictions 
of the closed-box model.
Then it becomes even more difficult
explaining the
observation as a global chemical enrichment. 
 An approximate way
of considering this open-box evolution 
using closed-box equations consists
of using effective yields inferred from
observations rather than 
the yield predicted by stellar
evolution models \citep[e.g.][]{dal07}.
These effective yields are smaller than 
true yields, and the difference increases
with decreasing galaxy mass (probably 
due to the decrease of the gravitational
potential of the galaxy). Actually,
the deviations are particularly large for 
dwarf galaxies, like our \qbcd s 
\citep[e.g.,][]{gar02,dal07}.
If the yield $y$ is reduced, then  
$Z-Z_s$ decreases too -- see
equation~(\ref{chem5}) and keep in
mind that $\mu$ is fixed by 
equation~(\ref{chem4}).
In short, the assumption of closed-box
chemical evolution do not
invalidate our conclusion, namely, that
the metallicity we infer from 
the emission lines traces only a small 
fraction of the total galactic gas.

So far we have put aside the behavior of the \bcd s. 
According  to Fig.~\ref{mainrc}, they have the same nebular
and stellar metallicities within error bars. 
In addition, the stellar
populations have ages slightly smaller than 1\,Gyr 
(Fig.~\ref{age}). 
These two properties can be easily explained if 
the \bcd\ galaxies are \qbcd s
experiencing a major but short starburst involving 
fresh well mixed galactic gas. In this case, the observed 
\bcd\ SFR is significantly larger than the 
average SFR during the age of the stellar 
population. 
Then the chemical evolution model to be applied 
has $M\sim M_*$, and a \sfr\  
of the order of the \sfr~of~\qbcd s,
i.e., a model similar to the case labelled 
as $2\,M_*$ in Fig.~\ref{deltaz}. The model prediction 
is $Z-Z_s$ at least one order of 
magnitude smaller than the difference observed in 
\qbcd s and, therefore, in agreement with the 
lack of metal enrichment observed in
\bcd s (the stars in Fig.~\ref{mainrc}).



\section{Conclusions}\label{conclusions}

We analyze the metallicity of \qbcd s, i.e., 
galaxies that may be Blue Compact Dwarfs (\bcd s) 
during the periods 
of quiescence, when the major starburst characteristic
of \bcd s is not so dominant. The \qbcd\ candidates 
were selected in \paperi\ from SDSS/DR6, 
where we also separate a reference sample of \bcd s. 
The metallicity inferred from emission lines 
of these \qbcd s turned out to exceed 
the metallicity  of the \bcd s, an uneasy result  
if \bcd s have to descend from \qbcd s. Here we 
study whether the metallicity inferred from the emission 
lines of \qbcd s may not be representative of the full
galactic gas, but reveals a local enrichment by 
recent starbursts. In this case the metallicities 
for the gas and for the stars must 
differ significantly.

The work is based on SDSS/DR6 spectra, whose
signal-to-noise ratio is not sufficient to 
measuring stellar metallicities from absorption lines
of individual spectra. We improve the original
signal-to-noise ratio by stacking observed
spectra that are alike. The grouping of similar 
spectra was carried out by classifying 
the 21493 \qbcd\  galaxies  using an automatic
{\em k-means} 
classification procedure (\S~\ref{class}). 
The algorithm renders a small number of 
types of spectra or {\em classes}, with the first 
ten classes containing 90\% of all spectra, and the 
most numerous Class~0 having more than 36\% of them. 
As a by-product, this classification scheme
provides  
a selective technique to 
identify galaxies in various states within 
the color sequence.  In particular, one of our 
classes  seems to contain only transition galaxies in 
the so-called {\em green valley} 
(Class~1, \S~\ref{green_valley}),
whereas another class includes most of the red 
sequence galaxies (Class~3; \S~\ref{green_valley}).
The typical Class~0 \qbcd\ galaxies belong
to the blue sequence. So do  
\bcd\ galaxies.

The stellar metallicities have been derived 
from the absorption lines using an ad-hoc 
procedure which fits the average profile of each class
with single-stellar populations
synthetic spectra based on the stellar
library MILES (\S~\ref{metalic}). 
We develop our own simple but robust tool 
to get a intuitive control of the errors.
Emission lines are masked out. 
The galactic continuum is also subtracted for fitting,
so that only absorption line features contribute to the
measurement.   
As inferred from our Monte-Carlo estimate of the
random error budget, a direct fit of the full spectrum
is good enough to assign ages, but it  does not provide 
enough finesse to properly distinguish metallicities.
Only after overweighting particular bandpasses of the 
stacked spectra \citep[corresponding to some  
of the Lick indexes defined by][]{wor94}, 
we bring the formal
error bars down to reasonable limits, 
below 0.1~dex.
Gas metallicities are obtained from  emission
lines with errors smaller than 0.2~dex
(\S~\ref{gas_metal} and App.~\ref{appa}).
When the gas and the stellar \qbcd\ metallicities are
compared, gas metallicities turn out to be systematically
larger than the stellar metallicities by some $\sim 0.35~$dex
(\S~\ref{metal_vs_metal}).
Despite the fact that this difference is not far from the
formal error bars (actually, it is below the formal 
2$\,\sigma$ level; \S~\ref{metal_vs_metal}), 
we regard it as significant for a number
of reasons. First, it is systematic, so that the
main \qbcd\ classes show it. Second, it is not present
in \bcd s, where stars and  gas show the
same metallicity within error bars  (Fig.~\ref{mainr}).
Third, the excess of gas metallicity with respect to
stellar metallicity is implicit in the 
luminosity-metallicity relationships for 
star-forming galaxies inferred by \citet[][gas]{tre04} and 
\citet[][stars]{gal05}
(see \S~\ref{metal_vs_metal}, and the dashed
line in Fig.~\ref{mainrc}).
Despite the existence of all these supportive 
arguments, a caveat is in order. The stellar 
metallicity error bars only describe statistical errors, 
although systematic errors may dominate the error budget.
Even if these systematic errors exist and are important, 
they should not modify 
the conclusions as they would affect both 
\qbcd s and \bcd s in the same way.
However, one can never discard a
source of (unknown and unsuspected)
systematic errors affecting \qbcd s and \bcd s 
differently, which would force us to 
reconsider the metallicity 
discrepancies.

The fraction of \qbcd\ galactic light produced by stars
augment with the metallicity, so that the fainter the emission
lines the more metal rich the gas. 
This result reinforces the conjecture that the emission
lines come from self enriched ISM. 
The luminosity weighted ages of \qbcd s span the full range
from 1 to 10 Gyr (Fig.~\ref{age}). The most common Class~0
is in the young part of such range, with an age below 2~Gyr.
The fact that the age of the stellar 
population of \bcd s is shorter but not very different 
from the age of \qbcd s adds on to this picture. 
If \bcd s and \qbcd s are
basically the same galaxies, but the \bcd s happen
to be in a phase of enhanced star formation activity, 
then the underlaying stellar populations must have 
similar properties. Their stellar metallicities 
are similar (\S~\ref{metal_vs_metal}).
The \bcd\ ages are smaller than the \qbcd s ages, 
but this can be 
easily due to the fact that their luminosity
weighted average ages are reduced by the current 
starburst.

In principle, the excess of metals in the ionized gas of
\qbcd s, as revealed by their emission lines, may 
reflect the natural enrichment of the ISM produced by 
successive SN ejecta. Emission lines trace the present 
ISM, whereas stars sample it in the past
when the metallicity was lower. The relative enrichment 
depends
on the age of the stars, and also
on the star formation rate providing the SNe. 
In the case of our \qbcd s,
these two quantities are tightly constrained. We have 
estimated the (mean) age of the starburst, and
the (current) star formation rate (SFR) as inferred from the 
observed H$\alpha$ luminosity \citep{ken98}. 
Using simple closed-box chemical evolution models, 
we argue that given the age and the star-formation 
rate, the observed starburst is not sufficient to 
enrich the full galactic ISM to the observed 
levels. However, age and SFR can be accommodated if the enriched 
galactic gas represents only a small fraction of the total
gas ($\sim$1/50; \S~\ref{chemical}). 
The assumption of closed-box 
evolution does not invalidate the conclusion. 

As we point out in \paperi , \qbcd\ are 
quiet common, representing one out each three 
dwarf galaxies in the local universe. 
Since they are so common, it is conceivable that some of 
their properties are not exclusive of the \qbcd\ class, 
but a global property of dwarf galaxies 
with emission lines. In particular, the bias of metallicity
inferred from emission lines may 
be present in all star-forming dwarf galaxies, 
rather than being a feature of our particular subset.
We plan to explore this potential 
bias using the techniques developed in the 
paper, namely, the comparison between the metallicity
estimates based on emission lines and absorption lines. 
Moreover, we plan of applying the classification tool in
\S~\ref{class} to find out and characterize  
galaxy spectra corresponding to the various parts
of the color sequence. The short green-valley phase
is particularly interesting
\citep[e.g., ][]{del07,sil08}, and we came across  
a simple method of identification.

%
%
%

\begin{acknowledgements}
Thanks are due to B.~Panter for clarifying discussions
on the proper solar abundance normalization to be used
with \citet{tre04} relationship.
Thanks are also due to an anonymous referee for 
helping us improving the argumentation. 
We benefitted from comments and suggestions by
R.~Amor\'\i n, I.~G. de~la~Rosa, C.~Esteban,
A.~Manpaso, 
M.~Moll\'a,
E.~P\'erez-Montero,
R. S\'anchez-Janssen, 
G.~Stasi\'nska,
and
J. V\'\i lchez
on aspects of this paper related to their area of 
expertise.
This work has been partly funded by the Spanish
{\em Ministerio de Educaci\'on y Ciencias}, projects
AYA~2007-67965-03-01 and AYA~2007-67752-C03-01.

    Funding for the Sloan Digital Sky Survey (SDSS) and SDSS-II has been provided by the Alfred P.
Sloan Foundation, the Participating Institutions, the National Science Foundation, the U.S. Department of Energy, the National Aeronautics and Space Administration, the Japanese Monbukagakusho, and the Max Planck Society, and the Higher Education Funding Council for England. The SDSS Web site is http://www.sdss.org/.
    The SDSS is managed by the Astrophysical Research Consortium (ARC) for the Participating 
Institutions. The Participating Institutions are the American Museum of Natural History, Astrophysical Institute Potsdam, University of Basel, University of Cambridge, Case Western Reserve University, The University of Chicago, Drexel University, Fermilab, the Institute for Advanced Study, the Japan Participation Group, The Johns Hopkins University, the Joint Institute for Nuclear Astrophysics, the Kavli Institute for Particle Astrophysics and Cosmology, the Korean Scientist Group, the Chinese Academy of Sciences (LAMOST), Los Alamos National Laboratory, the Max-Planck-Institute for Astronomy (MPIA), the Max-Planck-Institute for Astrophysics (MPA), New Mexico State University, Ohio State University, University of Pittsburgh, University of Portsmouth, Princeton University, the United States Naval Observatory, and the University of Washington.
\end{acknowledgements}

{\it Facilities:} \facility{Sloan (DR6, spectra)}



\appendix
 \section{Range of physical conditions where {\rm N2} holds}\label{appa}

\begin{figure}
\includegraphics[width=0.7\textwidth,angle=90]{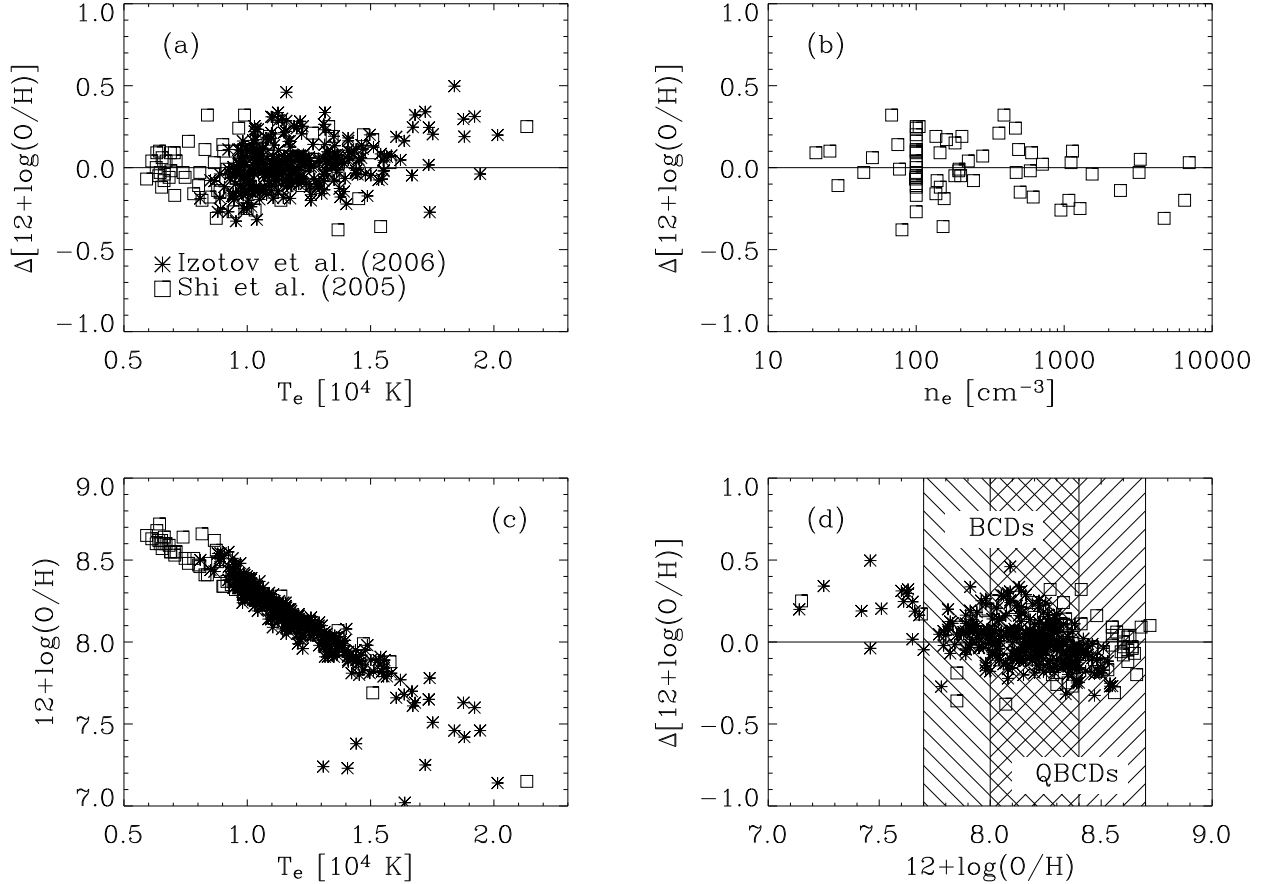}
\caption{(a) Difference between the oxygen abundances
determined with the N2 and the $T_e$~methods, 
$\Delta\large[12+\log({\rm O/H})\large]$,
for the set of galaxies analyzed by 
\citet[][the asterisks]{izo06} and 
\citet[][the squares]{shi05}.
They are represented vs the [OIII] electron temperature
$T_e$. There is no systematic trend for $T_e \la 17000$~K.
(b) $\Delta\large[12+\log({\rm O/H})\large]$
vs electron density $n_e$
for the set of galaxies analyzed by 
\citet[][]{shi05}.
No systematic trend.
(c) Tight correlation between
oxygen abundance and electron temperature
for $12+\log({\rm O/H}) \ga 7.5$.
(d) $\Delta\large[12+\log({\rm O/H})\large]$
versus  $12+\log({\rm O/H})$.
As indicated by the insets,
the hashed regions correspond to the range 
of abundances of \bcd s (the $-45$\degr\ slanted lines),
and the \qbcd s (the $+45$\degr\ slanted lines).
}
\label{calibra1}
\end{figure}

	The strong-line index abundance estimates are
calibrated against $T_e$ methods using particular sets of 
targets  \citep[e.g.,][]{dia00,pet04}. There is no guarantee that such 
calibrations hold  when applied to galaxies whose physical 
conditions deviate from those used for calibration.
In particular, if the physical conditions prevailing in 
the H~{\sc ii} nebulae  of \qbcd s and \bcd s 
are not included among the calibration targets,
then the metallicity difference between \bcd s and \qbcd s
found by \citet{san08} may be an artifact of using the 
strong-line N2 method.
The physical conditions in the  
\bcd\  H~{\sc ii} regions (newborn)
and in the  \qbcd s  H~{\sc ii} regions (aging) 
are expected to differ systematically which, coupled 
with an unfit N2~calibration, may give rise
to a false difference. 
Fortunately, such potential bias
seems to be harmless in our particular case. The N2 method
holds for a range of physical
conditions broad enough to encompass both \bcd s and 
\qbcd s.
Figure~\ref{calibra1}a shows the residual 
between the abundance determined with the 
N2~method, $\large[12+\log({\rm O/H})\large]_{\rm N2}$,
and the $T_e$ method,  $\large[12+\log({\rm O/H})\large]_{T_e}$,
\begin{equation}
\Delta\large[12+\log({\rm O/H})\large]=
\large[12+\log({\rm O/H})\large]_{\rm N2}-\large[12+\log({\rm O/H})\large]_{T_e},
\end{equation}
for the set of $\sim$310 metal-poor line-emission
SDSS/DR3 galaxies analyzed
by \citet{izo06}. The $T_e$ based abundances are directly 
computed by the authors from state-of-the-art photoionization
modeling, whereas the N2~abundances have been derived  from the
published line 
fluxes using  the recipe in \citet{pet04}.
\clearpage
The difference is represented as a function of the [OIII]
electron temperature. There is no obvious systematic 
difference between the two methods
when 5000~K $\le T_e \le $17000~K. The hint at systematic
deviation when  $T_e \ge $17000~K does not seem to affect
our estimates (see below). Figure~\ref{calibra1}a 
also includes the difference computed 
by \citet{shi05}\footnote{In this case the authors 
compute both the N2 and the $T_e$
abundance estimates. They differ systematically by 0.2 dex,
a constant term that we have removed when 
plotting since it does not affect relative differences 
between N2 abundances.} 
for a set of $\sim 70$ \bcd\ galaxies (the squares), which
we include because these authors also provide 
the second independent thermodynamic parameter,
namely, the electron density $n_e$.
Figure~\ref{calibra1}b shows the variation 
of the residual with $n_e$ for
\citet{shi05} galaxies, 
which evidences no systematic 
trend even though the large range
of electron densities that are involved;
 10~cm$^{-3}\le n_e\le 10^4$~cm$^{-3}$. 
Finally, Fig.~\ref{calibra1}d shows the residuals vs the
metallicity. Again residuals are independent
of metallicity except at the low end when
$12+\log({\rm O/H})\la\ 7.5$. 
The strong-line methods work because the seemingly
independent variables ($T_e$ and $n_e$) are actually related
to  $12+\log({\rm O/H})$ as a single parameter family 
\citep[e.g., ][\S~2.1.2]{sta04}. This is  the case for the 
dependence of $T_e$ on $12+\log({\rm O/H})$;
see  Fig.~\ref{calibra1}c. This tight correlation 
breaks down at  $12+\log({\rm O/H})\la\ 7.5$.
Due to the correlation,
the low abundance regime coincides
with the high temperature points which seems to deviate
from the nominal law as hinted in Fig.~\ref{calibra1}a.
This potential problem at low abundances and high temperatures
should not affect our sample of 
\bcd s and \qbcd s, whose 
abundances correspond to the well behaved part of the 
N2 calibration: see Fig.~\ref{calibra1}d,
where the range of \bcd\ and \qbcd\ abundances are 
represented as hashed regions. 
Moreover, according to Figs.~\ref{calibra1}a and \ref{calibra1}b,
the range of $T_e$ and $n_e$ where the N2 calibrations hold 
covers the full the range of thermodynamic parameters to be 
expected in H~{\sc ii} regions
\citep[e.g.,][]{dys00}. 
These two facts indicate that the N2 abundance 
estimate is valid for both \bcd s and \qbcd s, 
even though their H~{\sc ii} regions may have different 
physical conditions. Then the relative difference 
between the abundances of \bcd s and \qbcd s are
not  an artifact of using strong-line 
empirical methods. 
The {\em rms} variations of 
$\Delta\large[12+\log({\rm O/H})\large]$
are 0.16~dex and 0.13~dex, for \citet{shi05} galaxies
and \citet{izo06} galaxies, respectively.
In the case of \citet{izo06}, we only consider 
$12+\log({\rm O/H})\ge\ 7.5$.

\end{document}